\documentclass[conference]{IEEEtran}
\usepackage{cite}
\usepackage{amsmath,amssymb,amsfonts}
\usepackage{graphicx}
\usepackage{textcomp}
\usepackage{xcolor}
\usepackage{soul}
\usepackage{hyperref}
\usepackage{caption}% http://ctan.org/pkg/caption
\usepackage{subcaption}
\usepackage{cleveref}
\usepackage{multirow}

\def\BibTeX{{\rm B\kern-.05em{\sc i\kern-.025em b}\kern-.08em
    T\kern-.1667em\lower.7ex\hbox{E}\kern-.125emX}}

\usepackage{color}

\newcommand{\red}[1]{\textcolor{red}{\bf {#1}}}
\newcommand{\blue}[1]{\textcolor{blue}{\bf {#1}}}
\definecolor{darkspringgreen}{rgb}{0.09, 0.45, 0.27}

\begin{document}

\title{A Survey on Disaster: Understanding the After-effects of Super-cyclone Amphan\\ and Helping Hand of Social Media
}

\author{\IEEEauthorblockN{Soham Poddar\IEEEauthorrefmark{1}, Mainack Mondal\IEEEauthorrefmark{2} and Saptarshi Ghosh\IEEEauthorrefmark{3}}
\IEEEauthorblockA{\textit{Department of Computer Science \& Engineering} \\
\textit{Indian Institute of Technology, Kharagpur}\\
% sohampoddar [at] iitkgp.ac.in
% }
% \and
% \IEEEauthorblockN{Mainack Mondal}
% \IEEEauthorblockA{\small \textit{Dept. of Computer Science and Engineering} \\
% \textit{Indian Institute of Technology, Kharagpur}\\
% mainack [at] cse.iitkgp.ac.in}
% \and
% \IEEEauthorblockN{Saptarshi Ghosh}
% \IEEEauthorblockA{\small \textit{Dept. of Computer Science and Engineering} \\
% \textit{Indian Institute of Technology, Kharagpur}\\
% saptarshi [at] cse.iitkgp.ac.in}
Email: \IEEEauthorrefmark{1}sohampoddar [at] iitkgp.ac.in,
\{\IEEEauthorrefmark{2}mainack, 
\IEEEauthorrefmark{3}saptarshi\} [at] cse.iitkgp.ac.in
}
}

\maketitle

\begin{abstract}
The super-cyclonic storm ``Amphan'' hit Eastern India, specifically the state of West Bengal, Odisha and parts of Bangladesh in May 2020, and caused severe damage to the regions. In this study, we aim to understand the self-reported effects of this natural disaster on residents of the state of West Bengal. 
To that end, we conducted an online survey to understand the effects of the cyclone. In total, 201 participants (spanning five districts) from the worst-affected state of West Bengal participated in the survey. 
This report describes our findings from the survey, with respect to the damages caused by the cyclone, how it affected the population in various districts of West Bengal, and how prepared the authorities were in responding to the disaster. We found that the participants were most adversely affected in this disaster due to disruption of services like electricity, phone and internet (as opposed to uprooting of trees and water-logging). Furthermore, we found that receiving responses to Amphan-related  queries  is  highly  positively  correlated  with the  favorable  perception  of  people  about  preparedness  of authorities. 
Additionally, we study the usage of online social media by the affected population in the days immediately after the disaster. 
Our results strongly suggest how social media platforms can help authorities to better prepare for future disasters.
In summary, our study analyzes self-reported data collected from grassroots, and brings out several key insights that can help authorities deal better with disaster events in future.
\end{abstract}

\if 0
\fi 

\section{Introduction}
\label{introduction}

\noindent Super-cyclonic storm ``Amphan'' hit Eastern India, specifically the state of West Bengal, Odisha and parts of Bangladesh in May 2020. Amphan was a very powerful tropical cyclone -- technically, a super cyclone (CAT 5)~\cite{balasubramanian2020super} -- that caused widespread damage\footnote{\url{https://www.bbc.com/news/world-asia-52734259}}. In fact, CNN\footnote{\url{https://tinyurl.com/CNN-amphan}} reported that Amphan-caused damages amount to around US~\$13 billion, which makes it the costliest cyclone ever recorded in the North Indian Ocean. The cyclone made landfall in West Bengal on May 20, 2020.

The meteorological offices in India gave early warning of the development of Amphan. Consequently, state as well as central governments made an effort prior, during and after this natural disaster to contain the devastation. 
However, it is not well-understood how, from a  collective societal point of view Amphan affected the lives of \textit{normal} citizens, and if novel digital mediums like online social media platforms (e.g., Facebook, WhatsApp, Twitter) helped the citizens to cope better with this disaster. Thus, in this work we attempted to understand this collective view via self-reported data from a  sample of Amphan-affected population who resides in West Bengal, one of the worst affected states due to Amphan. 
Specifically, we conducted an {\it online survey} to get an estimate of the damages and impact of the cyclone on people. In this report we will present our insights gained from the survey.

Note that {\it cyclone Amphan occurred during the COVID19 pandemic, when West Bengal (and the whole of India) was under strict lockdown protocols}.  
This exceptional situation might have affected some of the observations in the survey. 
For instance, our survey indicates that people were much more adversely affected by factors such as disruption to phone and Internet services, than by physical damages such as uprooting of trees and waterlogging in their locality (as  detailed in later sections) -- the difference between the effects may have been larger due to the fact that the lockdown forced a lot of people to work from home, without venturing out into streets.
%Had the cyclone struck at normal times (when there would have been no restrictions on human mobility), these observations may have been different.
However, it is also possible that the authorities required more time to repair the damages (e.g., remove uprooted trees, repair damaged electricity cables) due to the ongoing pandemic (and resultant lockdown), than they would have under normal circumstances.
Also note that, our online mode of survey deployment was the only feasible option during a nation-wide lockdown (as opposed to approaching citizens offline).

\subsection{Survey methodology}

\noindent We deployed the survey on May 30, 2020, i.e., 10 days after the cyclone. 
We chose to wait for 10 days after the cyclone due to several reasons -- (i)~the Internet connectivity of many people was disrupted by the cyclone for several days (as will be evident from the results of the survey discussed in later sections). 
Since our survey would be conducted online, we wanted to ensure that Internet connectivity would be restored for most people, and 
(ii)~to allow people time to recuperate from the effects of the cyclone. 

The survey was disseminated through mailing lists of a few educational institutes in and around Kolkata, West Bengal, and through social media posts by the authors. Consequently, as can be seen in Section~\ref{demographics}, most of the participants of the survey were students and faculty members of educational institutes in and around Kolkata. 
We collected data using our survey over a duration of $2$ weeks, i.e., between May 30 and June 16, 2020. In total, 201 participants responded to the survey.

As part of the survey, a participant was mostly asked about the effects of the cyclone in his/her `locality' which was described as the region approximately within 500 metres in all directions from the participant's home.\footnote{The exact questions asked during the survey are listed in the Appendix.} 
Among the participants, 90.6\% said they were themselves in the locality for which they were answering during the cyclone, while the rest were responding on behalf of others who were in affected localities.

\subsection{This report}

\noindent This report analyses the responses to the survey.
We start with the demographics of the participants in Section~\ref{demographics}. 
We then get an estimate of the damages caused by the cyclone to various localities and districts (Section~\ref{damages}),
followed by how much the damages affected the people (Section~\ref{affected}). 
%We did some further analysis on effects of disruption Electricity, Phone and Internet services (which seem to be the utilities affecting people the most) in Section~\ref{utilities}. 
Then we discuss the use of Online Social Media (OSM) by the people affected by the cyclone (Section~\ref{socialmedia}). 
We also analyse how well the authorities were prepared to deal with the crisis (Section~\ref{authorities}).
%Section~\ref{authorities} discusses the preparedness of authorities to deal with the disaster.
Further, we discuss some implications/takeaways from the survey that can help better handling of disasters in future (Section~\ref{sec:implications}).
We conclude the study in Section~\ref{sec:conclu}.

%The contributions of this report can be summarised into the following -- We get an estimate of the damages caused by the cyclone, and how much people were affected by them. We then analyse the effects of social media on people after the disaster, if it helped spread important information. 

\vspace{4mm} 
\noindent {\bf Key takeaways:} Some of the key takeaways from the survey are listed below. Details can be found in later sections.
\begin{itemize}
    \item People were most affected due to disruption of services such as electricity, phone and internet, than due to physical damages like uprooting of trees (this observation might have been affected by (i) the COVID19-induced lockdown which meant most people were indoors and (ii) the bias in our survey participants towards students and academicians residing in urban areas).
    \item Among the two most popular electric suppliers in West Bengal, it seems that CESC managed to restore a significantly higher fraction of connections (over 60\%) within a day after the cyclone, as compared to WBSEDCL. However, for both suppliers, there were a comparable fraction of hard-to-repair connections whose restoration took more than $4$ days.
    \item Among the popular phone service providers, Jio seems to have performed the best, with about 52\% of the connections being disrupted, compared to almost 89\% of Vodafone connections and 79\% of Airtel connections being disrupted.
    \item The districts of Howrah and South 24 Parganas seem to have been the most affected, with very high fractions of participants saying they were severely affected by multiple factors associated with the cyclone. Though, it is to be noted that our sample sizes from these two districts are relatively small.
    \item It is seen that Online Social Media (OSM) is being used by the affected population in several important ways, including inquiring about safety of others, informing others of one's own safety, organizing donation campaigns, and so on. However, the authorities are {\it not} using OSM sufficiently to connect to people, which should be considered by the authorities for the future. There will always be the risk of fake news/biased opinions spreading through OSM (as also observed to some extent after Amphan), but the benefits of mass communication through social media clearly outweigh the downsides.
    \item Participants have given mixed ratings to the preparedness of the authorities in dealing with the disaster -- the number of people who opined that the authorities were highly prepared to deal with the disaster almost equals the number of people who felt that the authorities were poorly prepared. We observed that people tend to give higher ratings to the authorities if they received helpful responses from the authorities, and if their services were restored earlier.
\end{itemize}
\section{Demographics of Participants}
\label{demographics}

\noindent In this section, we study the demographics of the 201 participants who responded to the survey, in terms of their gender distribution, age distribution and occupation distribution. 
We also compare the distribution of the participants with the overall population distribution of the state of West Bengal, to check how representative our set of participants is of the overall population of the state.

\vspace{2mm}

\noindent \textbf{Gender distribution:}
Table~\ref{tab2:gender} shows the distribution of gender of the survey participants\footnote{We also had `Others' as an option (apart from `Male' and `Female') in our survey. However, no participant selected it.}.
We also compare it to the distribution of gender in the overall population of the state of West Bengal according to 2011 census 2011\footnote{\url{https://www.census2011.co.in/census/state/west+bengal.html}}. 
As we can see, the gender ratio of the survey participants is slightly biased towards males, with males consisting of 58\% of the participants compared to 51\% in overall population of West Bengal. 
%However, this bias is also present in the overall population  of West Bengal. 
%Gender Ratio is a bit biased towards male but mostly in accordance to the that in WB. 
 
\begin{table}[hbt]
    \centering
    \normalsize
    \begin{tabular}{|l|c|c|}
        \hline
         & \textbf{Male} & \textbf{Female}\\
        \hline
        \textbf{Survey Participants} & 58.2\% & 41.8\% \\
        \textbf{Population of West Bengal (WB)} & 51.3\% & 48.7\% \\
        \hline
    \end{tabular}
    \caption{\textbf{Gender distribution of survey participants and people in West Bengal. The gender ratio of the participants is slightly biased towards males.}}
    \label{tab2:gender}
\end{table}

\begin{figure}[hbt]
    \centering
    \includegraphics[width =0.8\linewidth,height=5cm]{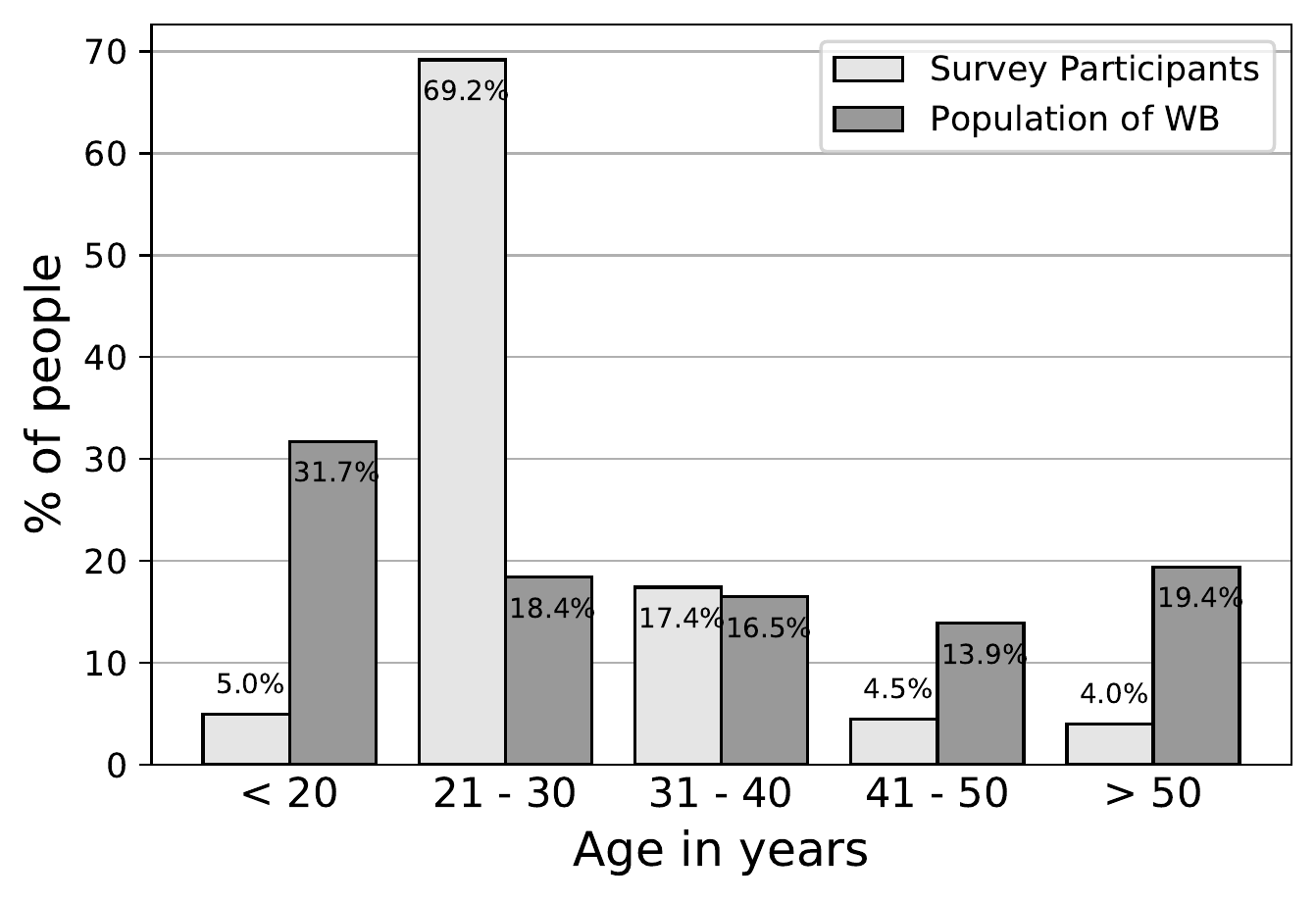}
    \caption{\bf Age distribution of survey participants and the overall population of West Bengal (WB). Our sample is heavily biased towards the age group between 21-40.}
    \label{fig2:age}
\end{figure}

\vspace{2mm}
\noindent \textbf{Age distribution:} Figure~\ref{fig2:age} presents the age-distribution of our participants. We again compare the age distribution to that of the overall population of West Bengal according to the  2011 census\footnote{\url{http://statisticstimes.com/demographics/population-of-west-bengal.php}}. 

Evidently, the age distribution of our participants is heavily biased towards people aged between 21 and 30 years (69\% among survey participants, compared to 18\% in WB population). 
In fact, 85\% of the survey participants are aged between 21 and 40 years (compared to 35\% in WB). 
This bias towards younger participants is mainly a result of  our methodology of distributing the survey via mailing lists of educational institutions, whereby the survey was mostly taken by university students. 
However, this might even be beneficial for our study since people in this age group are generally well aware of the situation in the localities and their families. 
In fact, 90\% of the participants in this age-group were themselves in the localities for which they reported the damages.

\begin{figure}[hbt]
    \centering
    \includegraphics[width =0.8\linewidth,height=5cm]{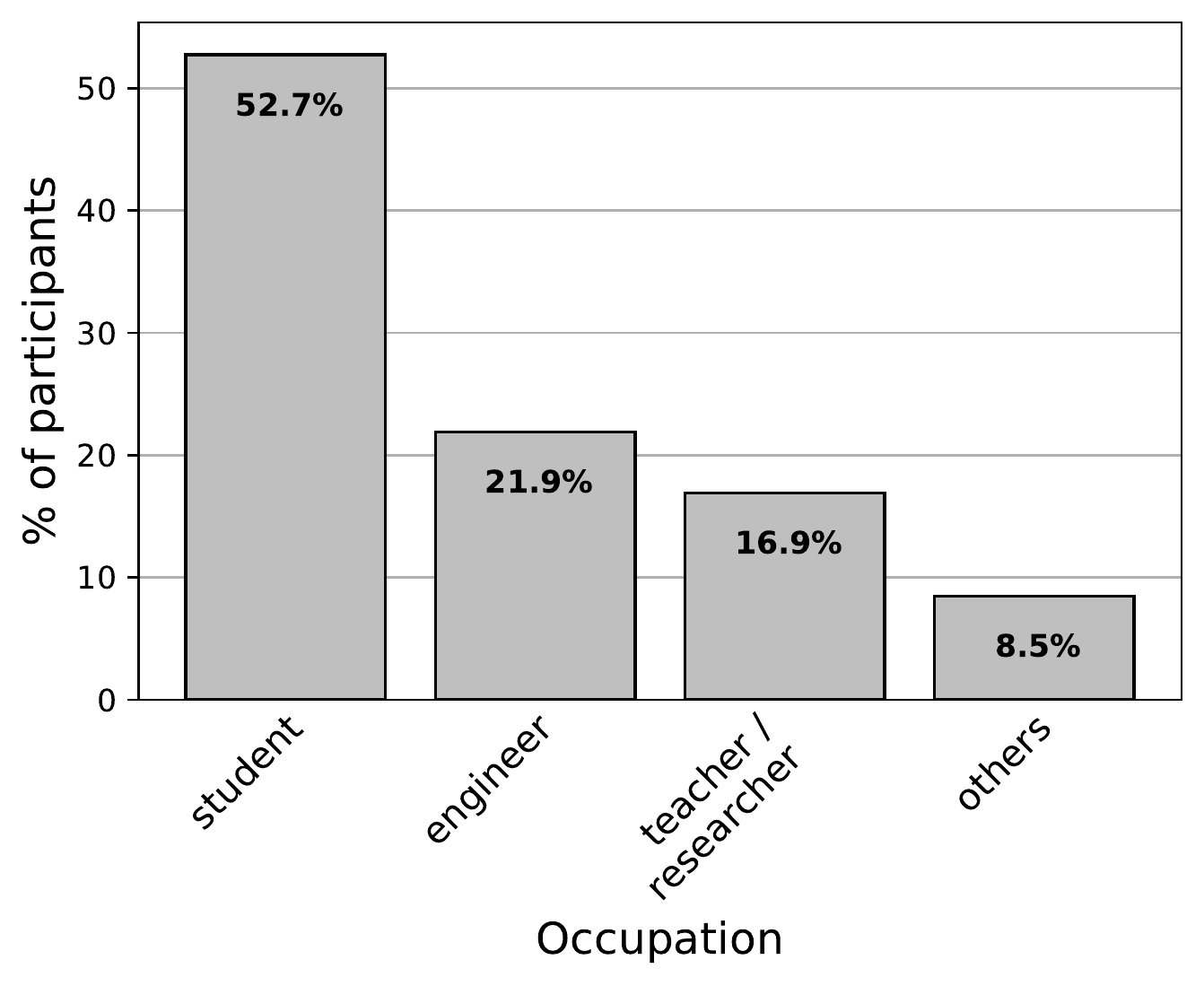}
    \caption{\bf Occupation distribution of survey participants. The participants are mostly students.
    }
    \label{fig2:occupation}
\end{figure}

\vspace{2mm}
\noindent \textbf{Occupation distribution:} The distribution of occupation of the participants is given in Figure~\ref{fig2:occupation}. Our sample is biased towards more students, teachers and engineers, with students making up about 53\% of our sample. 
As in the case of age distribution, this bias stemmed from our recruitment procedure which relied heavily on the students and faculty members of educational institutes, and users of social media, who are more likely to use the Internet (and take our survey).

\begin{figure*}[hbt]
\begin{subfigure}{0.5\textwidth}
    \centering
    \includegraphics[width =0.9\linewidth,height=6cm]{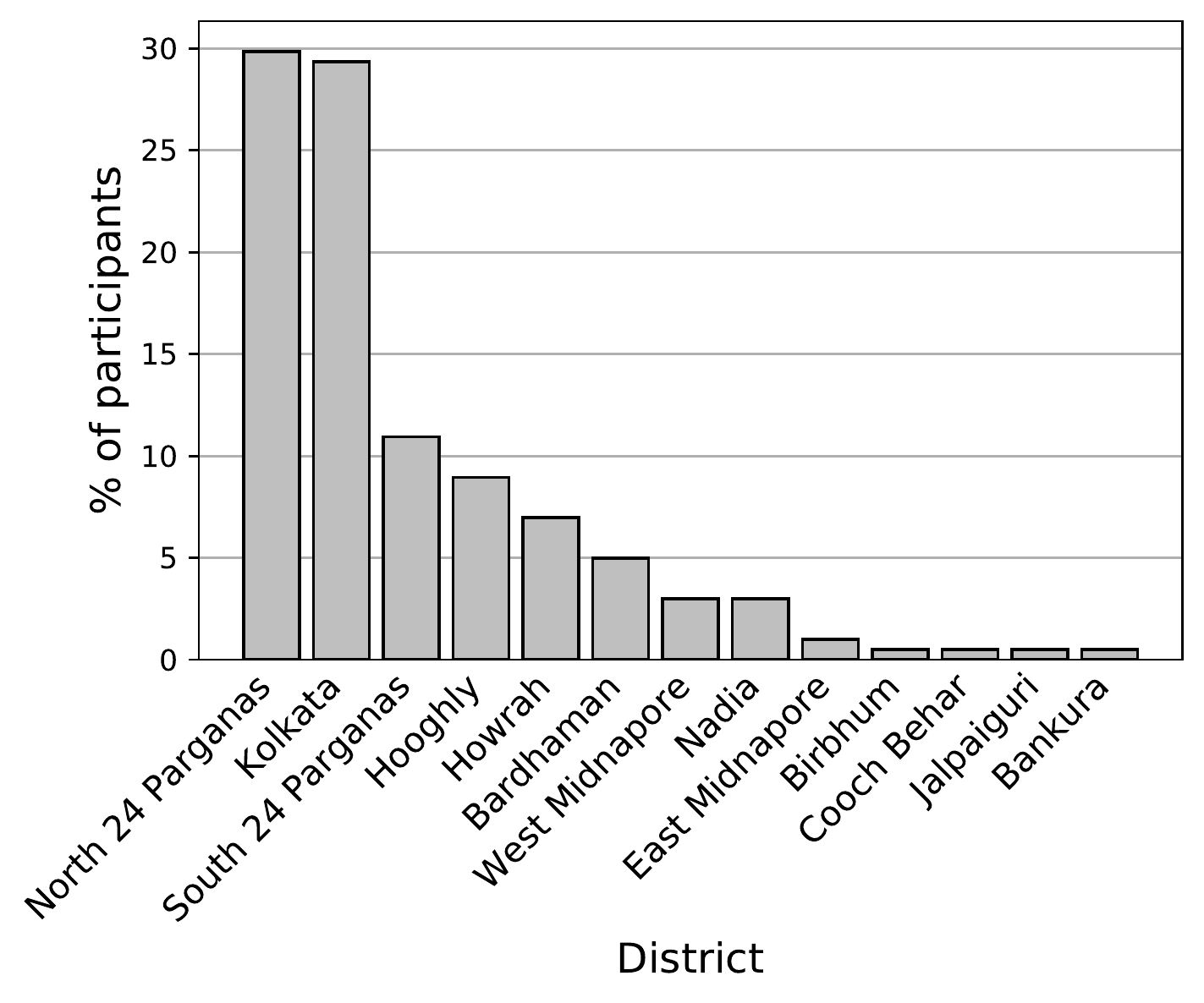}
    \caption{\bf Distribution of participants among all districts in WB.}
    \label{fig2:district_all}
\end{subfigure} % 
\hfill
\begin{subfigure}{0.5\textwidth}
    \centering
    \includegraphics[width =0.9\linewidth,height=6cm]{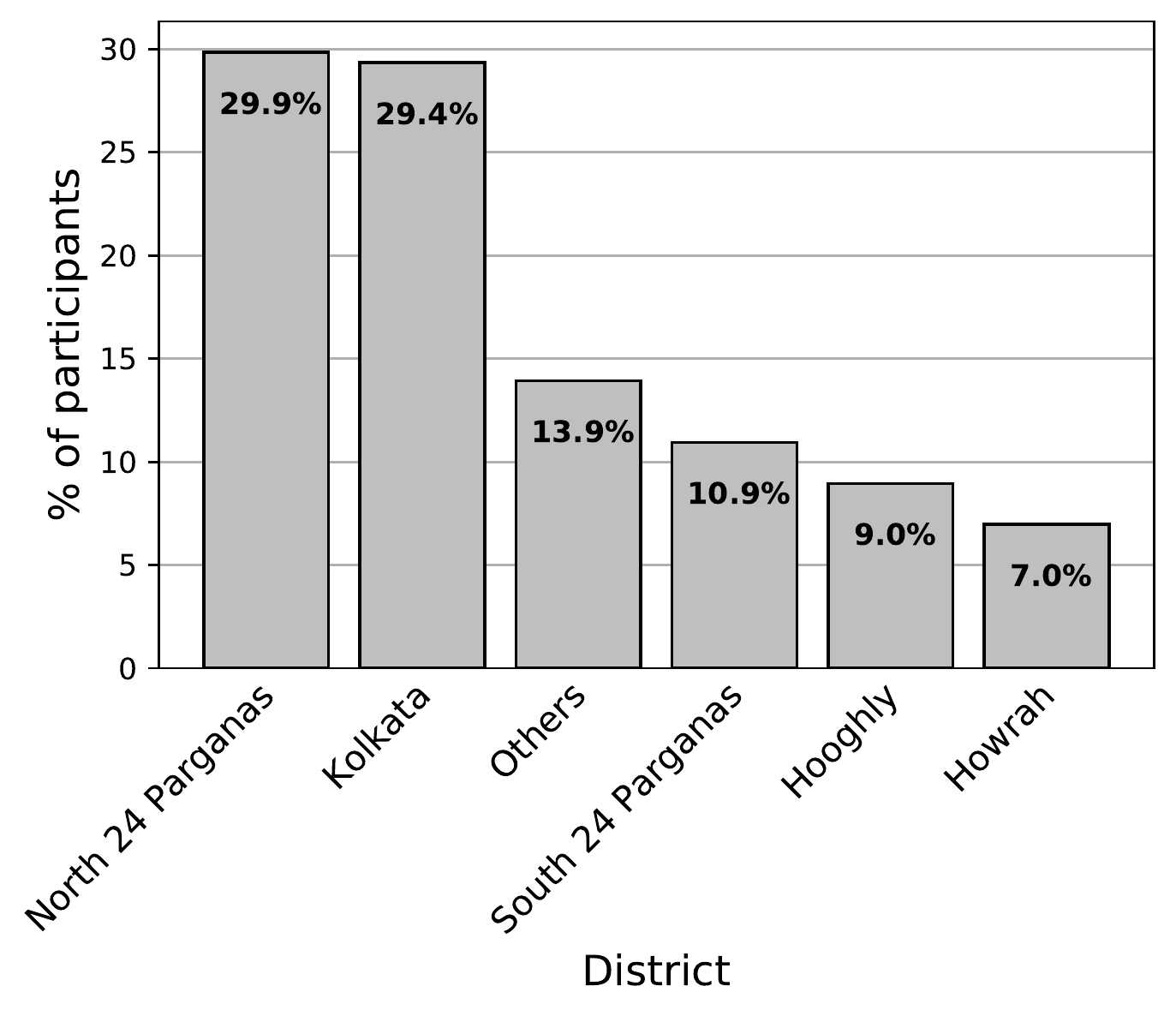}
    \caption{\bf Considering the top 5 districts and grouped rest as ``Others''.}
    \label{fig2:district5}
\end{subfigure}
\caption{\bf Geographical distribution of survey participants. Most of the participants were from N24P and KOL.}
\label{fig2:districts}
\end{figure*}

\vspace{2mm}
\noindent \textbf{Geographical distribution}: 
During the survey, participants were asked to state the pin code of their locality.  
Responses were obtained from a total of 124 distinct pin codes.
We mapped the pin codes to the various districts of West Bengal. The distribution of the pin codes among the districts of West Bengal are as shown in Figure~\ref{fig2:district_all}. 
According to the Wikipedia\footnote{\url{https://en.wikipedia.org/wiki/Cyclone_Amphan}}, and the news paper ``Hindustan Times''\footnote{\url{https://tinyurl.com/HT-amphan}}, the districts North 24 Paraganas (N24P), South 24 Parganas (S24P), Kolkata (KOL), Hooghly (HGLY) and Howrah (HWH) were the ones mostly affected by cyclone Amphan. These are also the five districts from where most of the responses were received. 

We have considered these top 5 districts individually and grouped the rest as ``Others'', as in Figure~\ref{fig2:district5}. 
It can be noted that, although the majority of the survey participants were from urban areas, some of them were from semi-urban areas.

\subsection*{Summary of Section}
\noindent We found that our sample population is biased towards a younger population (age group of 21-40), and towards students. 
This bias is inherent in our recruitment procedure (via email list of educational institutions, social media). 

However, we note that this biased population is still suitable for our study due to two reasons: 
First, our sample covers a wide range of cyclone-affected localities and provides a wide coverage over age groups as well as gender.
Moreover, this bias might even facilitate parts of our exploration -- one of the aims of this study is to understand the perceptions about digital media usage during disasters; since our survey respondents are well-versed with digital media usage, it is possible to gain valuable insights about this aspect from the survey (see Section~\ref{socialmedia}).

\section{Measuring Extent of Damages}
\label{damages}

\noindent In this section, we study the damages caused by the cyclone to the various localities of the participants. We first analyse the physical damages caused to the localities (for example, damages to trees and buildings) in Section~\ref{physicaldamage}, followed by an analysis of how services like electricity supply were disrupted in Section~\ref{utilities}.

\subsection{Physical damages to localities}
\label{physicaldamage}

\begin{figure*}[hbt]
\begin{subfigure}{0.5\textwidth}
    \centering
    \includegraphics[width =0.9\linewidth,height=5cm] {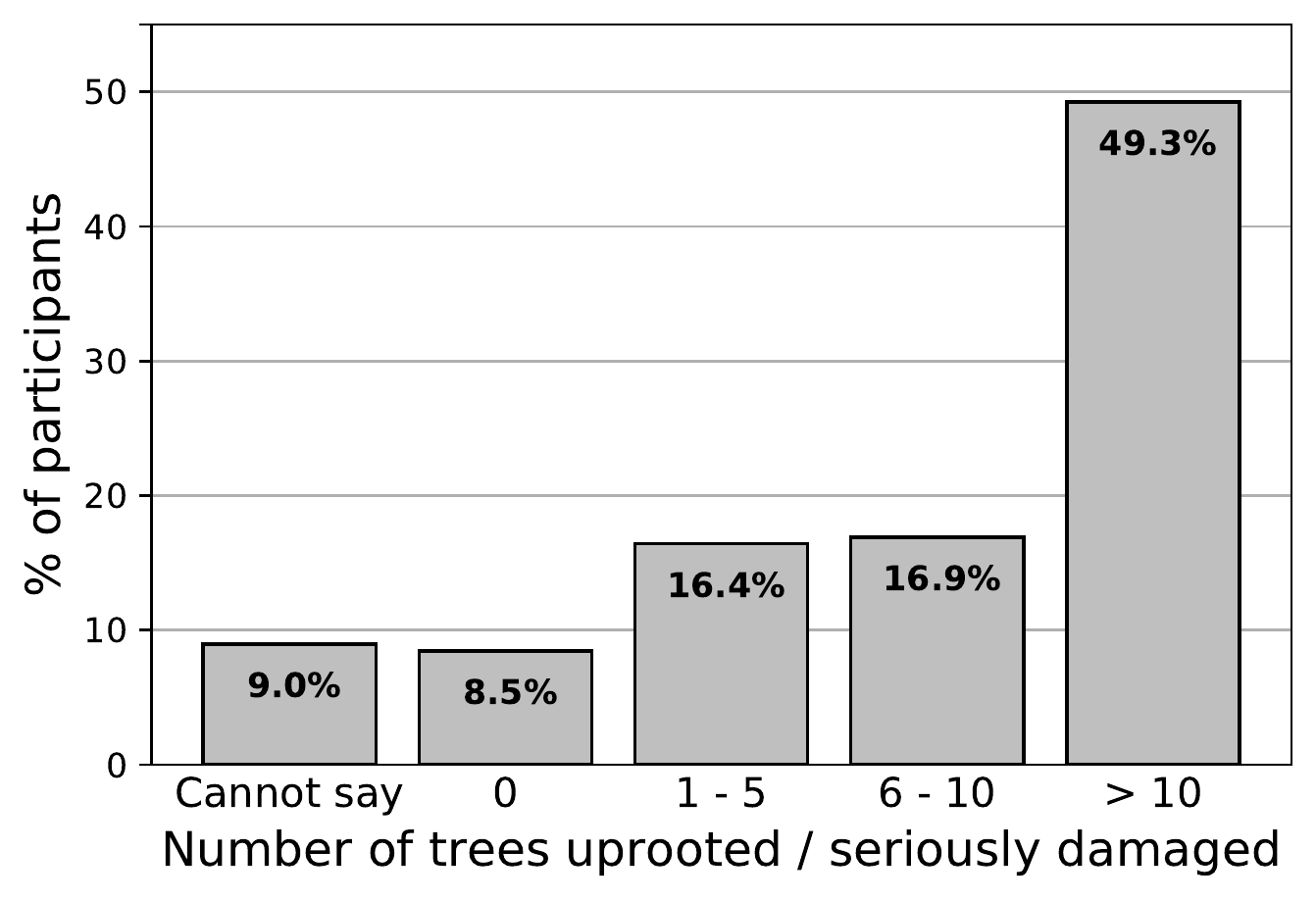}
    \caption{\bf \#Trees uprooted in participants' locality.}
    \label{fig3:treedamage}
\end{subfigure}%
\hfill
\begin{subfigure}{0.5\textwidth}
    \centering
    \includegraphics[width =0.9\linewidth,height=5cm]{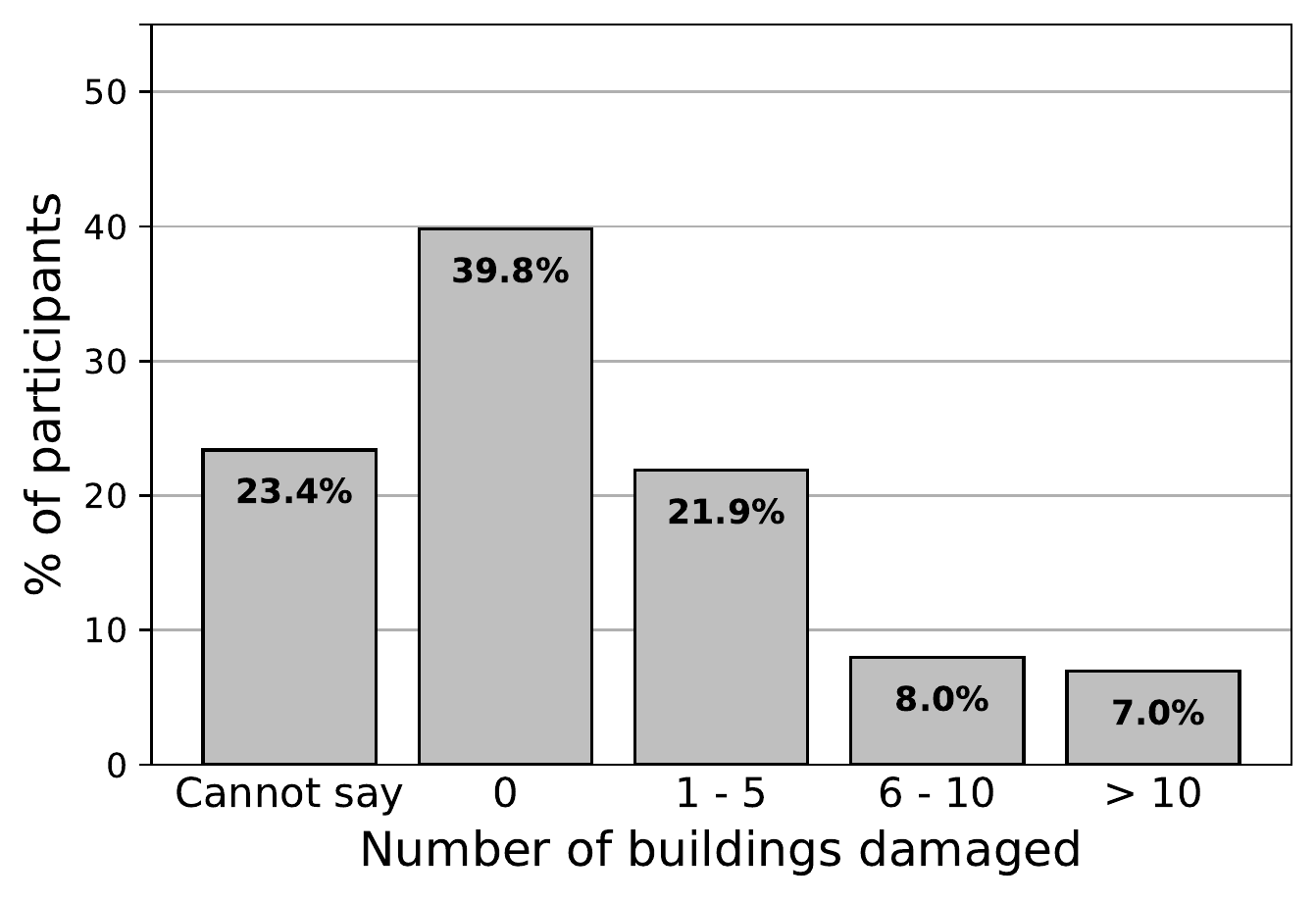}
    \caption{\bf \#Buildings damaged in participants' locality.}
    \label{fig3:buildingdamage}
\end{subfigure}
\newline
\begin{subfigure}{0.5\textwidth}
    \centering
    \includegraphics[width =0.9\linewidth,height=5cm]{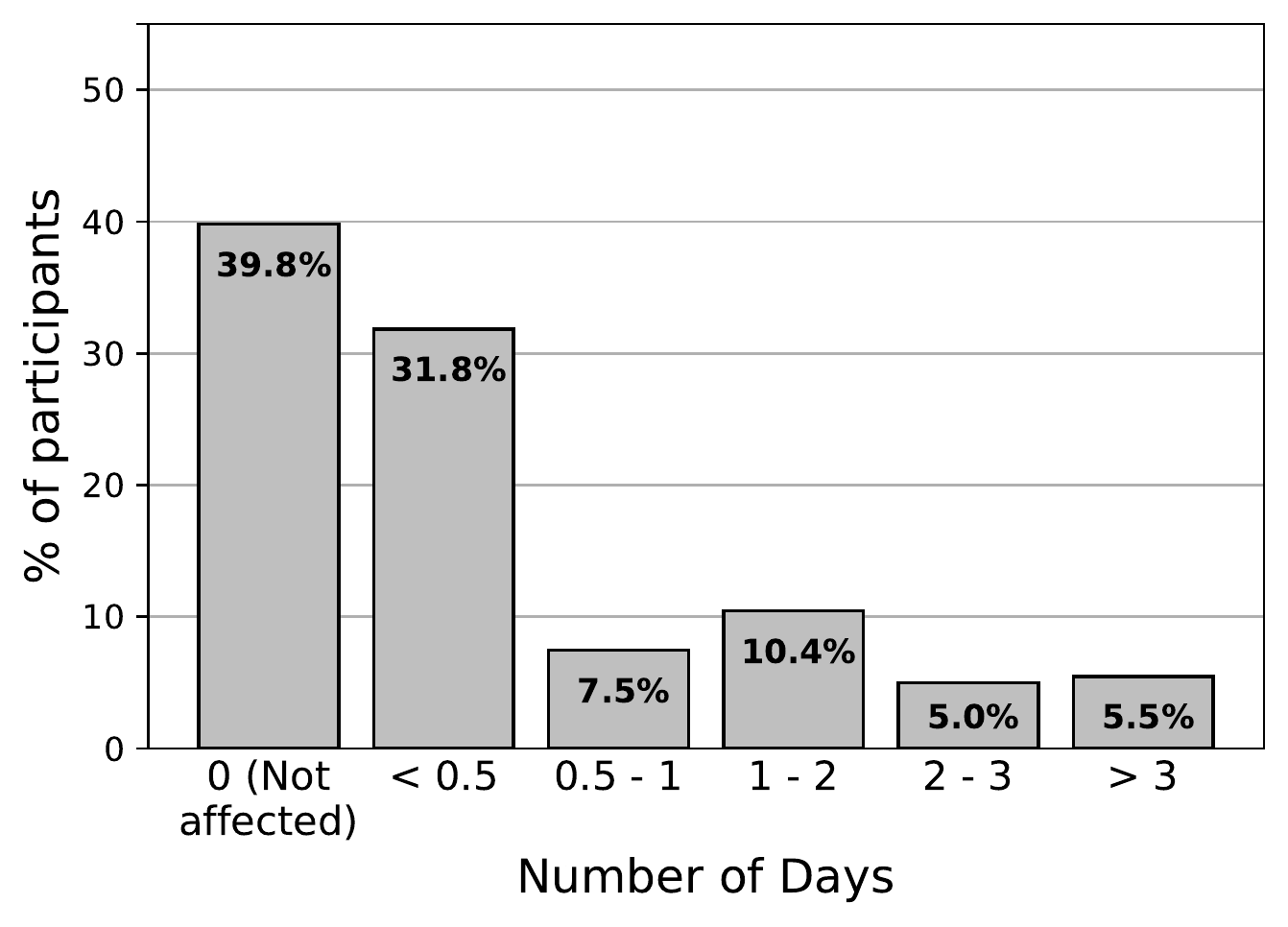}
    \caption{\bf \#Days for which participants' locality was waterlogged.}
    \label{fig3:waterdamage}
\end{subfigure}%
\hfill
\begin{subfigure}{0.5\textwidth}
    \centering
    \includegraphics[width =0.9\linewidth,height=5cm]{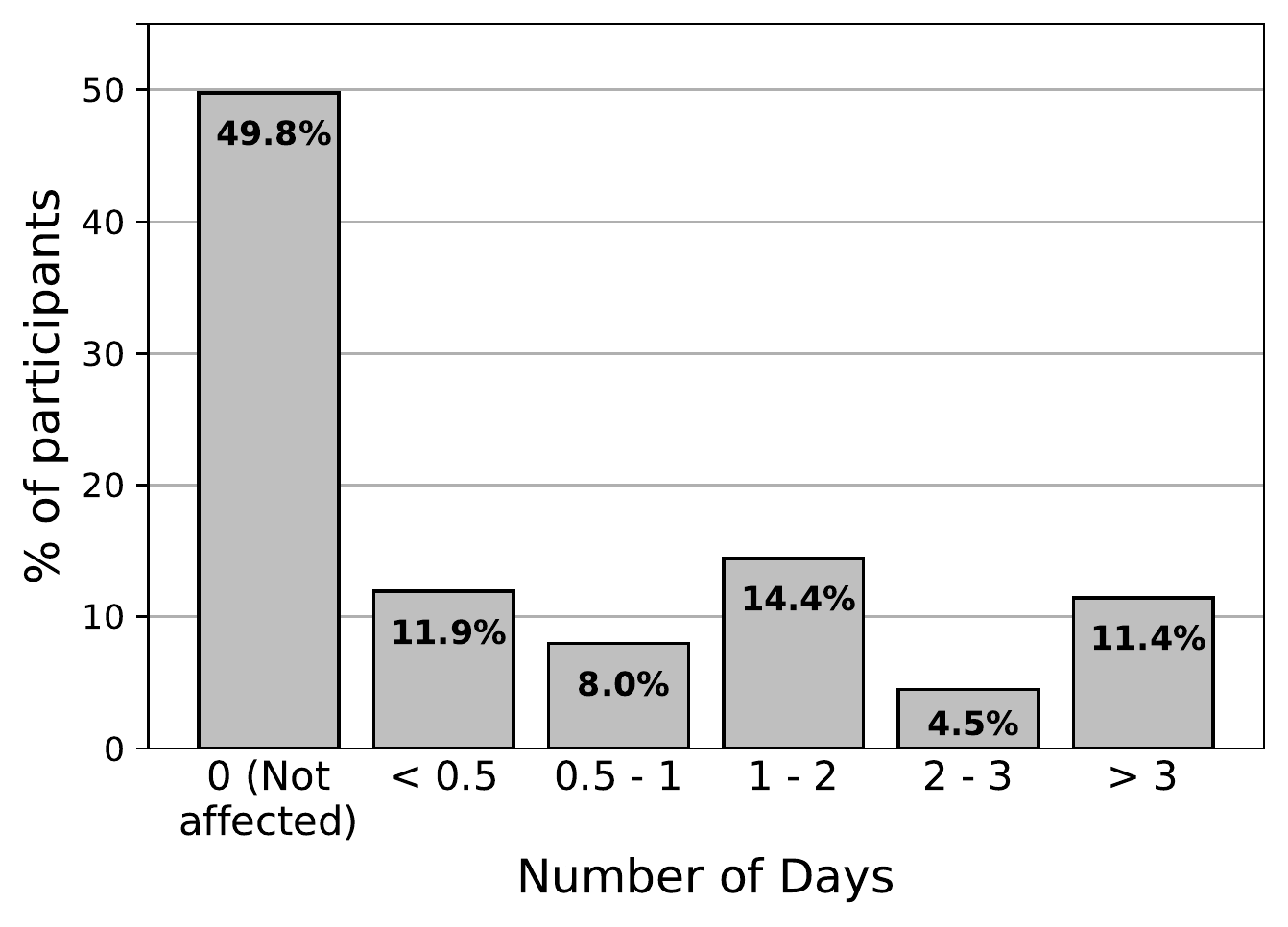}
    \caption{\bf \#Days for which participants' locality faced disruption of supply of drinking water.}
    \label{fig3:drinccdamage}
\end{subfigure}

\caption{\bf Measuring the extent of damages in the localities of the survey participants. A lot of trees were damaged in many localities, but there was not much damage to buildings in the localities of the participants (mostly urban). Waterlogging and disruption to drinking water supply have also been relatively less.}
\label{fig3:alldamages}
\end{figure*}

\noindent We asked participants to give an estimate of the damages caused by Amphan to their localities in terms of (i)~number of trees uprooted, (ii)~number of buildings damaged, (iii)~number of days for which one's locality was waterlogged. 
We had specified that ``locality'' refers to the region approximately within 500 meters in all directions from the participant's home. 
The distribution of responses by the participants are shown in Figure~\ref{fig3:alldamages}. 
We have also calculated the district-wise distribution of heavily damaged localities, and tabulated them in Table~\ref{tab3:districtdamage}. 

\vspace{2mm}
\noindent \textbf{Trees uprooted:} The number of trees that were uprooted or seriously damaged by the cyclone is shown in Figure~\ref{fig3:treedamage}. Uprooting of trees was prevalent in a large majority of localities. In 66.2\% of the localities, more than 5 trees were uprooted. In about 49.3\% localities, more than 10 trees were uprooted. 

The district wise distribution of heavy damage to trees is given in the first row of Table~\ref{tab3:districtdamage}. As expected, damages to trees affected the other districts (``Others'') quite less (28.6\% participants reported more than 5 trees uprooted), whereas for each of the 5 districts that were most affected by the cyclone, over  65\% participants reported more than 5 trees uprooted. The fractions are especially high in South 24 Paraganas, Hooghly and Howrah, where more than 85\% participants from these districts reported more than 5 trees uprooted in their locality.
%The possible reason for this may be that the districts in “Others” group  have been affected much less by Amphan. South 24 Paraganas has a lot of trees and hence it is expected to have more uprooted trees.

\vspace{2mm}
\noindent \textbf{Buildings damaged:} The number of buildings damaged in Figure~\ref{fig3:buildingdamage}. Damages to buildings was relatively less in the localities of our survey participants. Only in 15\% of the localities, more than 5 buildings were damaged. From Table~\ref{tab3:districtdamage} we see that the fraction of participants who reported more than 5 buildings damaged is 10\% in North 24 Paraganas, 13.6\% in Kolkata, slightly more (27.8\%) in Hooghly. 
These low fractions are somewhat expected due to these being urban areas.

\vspace{2mm}
\noindent \textbf{Waterlogging:} The distribution of the duration for which the localities were waterlogged is shown in Figure~\ref{fig3:waterdamage}. About 21\% of all localities of the participants faced waterlogging issues for more than 1 day. As seen in Table~\ref{tab3:districtdamage} waterlogging affected Howrah the most (35.7\% participants reported waterlogging for more than one day) and Hooghly the least (5.6\% participants reported waterlogging for more than one day). 

The relatively low severity of waterlogging in some localities indicates good drainage facilities in those localities. The areas where waterlogging was more severe could have improper drainage that tend to flood even with normal rains. 
In fact, 31.2\% of the survey participants said that their locality was waterlogged due to cyclone Amphan, but is not usually waterlogged during normal rainfall. 
Whereas, 23.8\% participants said that their locality gets regularly flooded every time there is heavy rainfall. The authorities should attempt to improve the drainage facilities in these localities.

\begin{table*}[hbt]
    \centering
    \normalsize
    \begin{tabular}{|l||c|c|c|c|c||c|}
        \hline
        \shortstack[l]{\bf Participants who said their locality was damaged\\ \bf heavily in terms of:} & \shortstack{\bf N24P\\(60)} & \shortstack{\bf KOL\\(59)} & \shortstack{\bf S24P\\(22)} & \shortstack{\bf HGLY\\(18)} & \shortstack{\bf HWH\\(14)} & \shortstack{\bf Others\\(28)} \\
        \hline\hline
        Trees uprooted (More than 5 trees) & 65.0\% & 66.1\% & 86.4\% & \textbf{88.9\%} & 85.7\% & 28.6\% \\
        
        Buildings damaged (More than 5 buildings) & 10.0\% & 13.6\% & 18.2\% & \textbf{27.8\%} & 21.4\% & 14.3\% \\
        
        Waterlogging (More than 1 day) & 21.7\% & 27.1\% & 18.2\% & 5.6\% & \textbf{35.7\%} & 10.7\% \\
        
        Drinking water supply disrupted (More than 1 day) & 35.0\% & 22.0\% & 27.0\% & \textbf{44.4\%} & 42.9\% & 21.4\% \\
        
        %\hline
        
        Electricity supply disrupted (More than 1 day) & 56.7\% & 25.4\% & {\bf 81.8\%} & 61.1\% & 64.3\% & 28.6\% \\
        Phone service disrupted (More than 1 day) & 61.7\% & 72.9\% & {\bf 81.8\%} & 61.1\% & 64.3\% & 25\% \\
        Internet service disrupted (More than 1 day) & 68.3\% & 88.1\% & 81.8\% & 83.3\% & {\bf 100\%} & 25\% \\
        
        \hline
    \end{tabular}
    \caption{\bf Distribution of heavy damages to different districts, where `heavy damage' is as described in the first column. The percentages represent the fraction of participants from that district who said their localities were damaged heavily. The numbers in brackets below each of the district names represent the count of participants from that district. The highest percentage in each row is highlighted in boldface. Note that ``Others'' refers to all the districts that were {\it not} affected much by cyclone Amphan.}
    \label{tab3:districtdamage}
\end{table*}

\subsection{Disruption of services}
\label{utilities}

\noindent We asked the participants if their drinking water, electricity, phone and internet services had been disrupted due to the cyclone Amphan, and if so, how long it took for these services to be restored back. 
We now analyse the disruption of various services.

\vspace{2mm}
\noindent \textbf{Disruption of drinking water supply:} The duration for which the supply of drinking water was disrupted is given in Figure~\ref{fig3:drinccdamage}. 
About 30.3\% of the participants faced problems with drinking water supply beyond one day. The district wise distribution is given in the Table~\ref{tab3:districtdamage}. Drinking water problems were faced by all districts more or less evenly, with Hooghly and Howrah facing the most problems, with 44.4\% and 42.9\% of the participants from these districts saying they faced irregular drinking water supply beyond one day. 

It can be noted that disruption of drinking water supply may be correlated with electricity services being hampered for a long time (as discussed below) which  rendered electric water pumps inoperable.

\vspace{2mm}
\noindent \textbf{Disruption of electricity supply:} 
There are two primary electricity supply providers in West Bengal -- (i)~CESC that primarily supplies Kolkata and parts of Howrah and other districts neighboring Kolkata, and (ii)~WBSEB/WBSEDCL that supplies the other regions.
In total 91 of our survey participants reported they were supplied by CESC, and 106 participants reported they were supplied by WBSEDCL/WBSEB (while only 4 reported some other supplier).

In total, as many as 89.1\% of our survey participants reported disruption of electricity in their locality.
Specifically, 82.4\% out of the 91 participants supplied by CESC and 95.3\% out of the 106 participants supplied by WBSEDCL/WBSEB faced disruption of electricity supply.
Thus, it seems that people supplied by WBSEDCL/WBSEB were disrupted more than those by CESC (95.3\% vs 82.4\%).

\begin{figure*}[hbt]
\begin{subfigure}{0.45\textwidth}
    \centering
    \includegraphics[width =\linewidth,height=5cm]{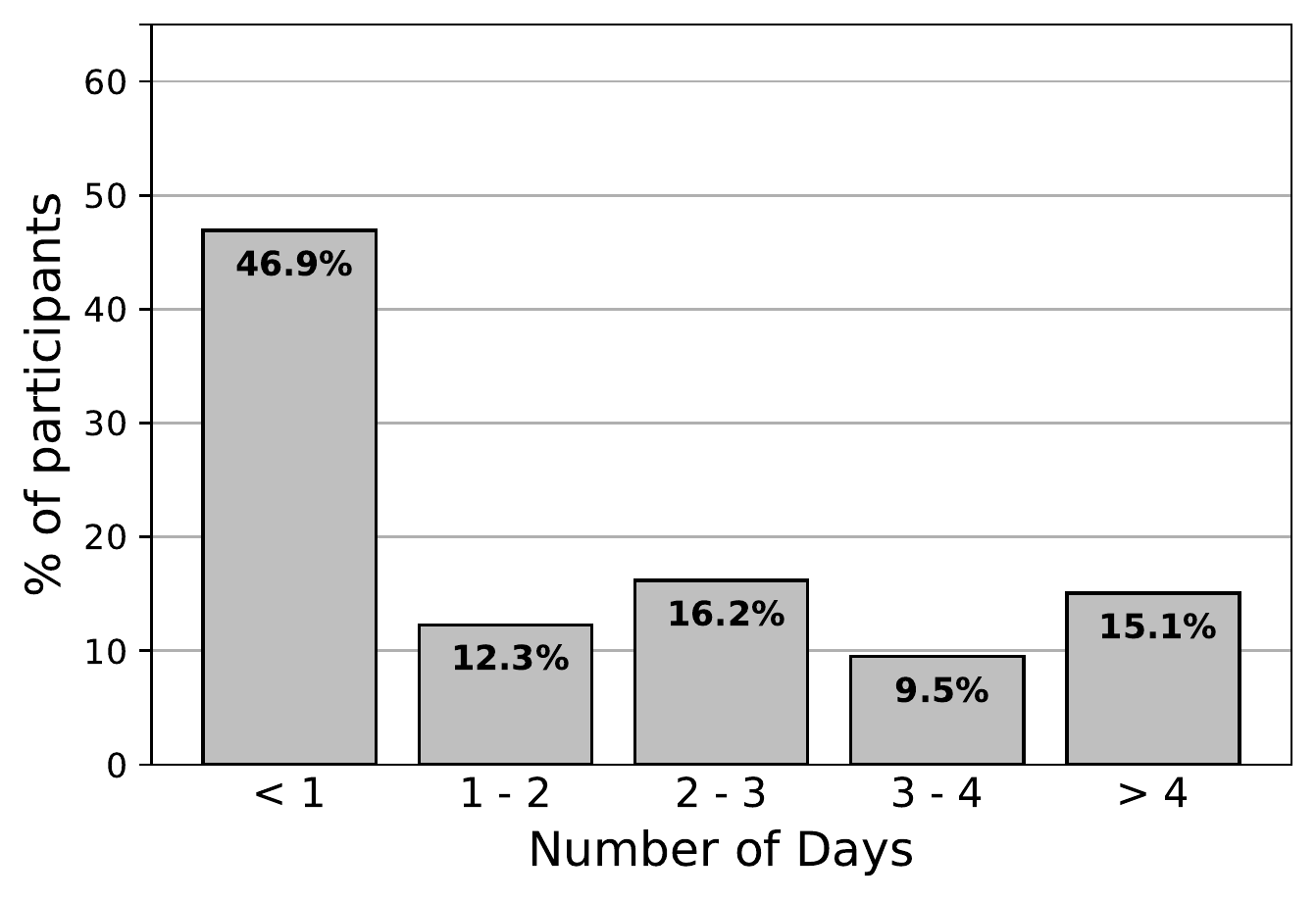}
    \caption{\bf \# Days taken for restoration of electricity (for all 201 participants).}
    \label{fig5:electricrestore}
\end{subfigure}%
\hfill
\begin{subfigure}{0.45\textwidth}
    \centering
    \includegraphics[width =\linewidth,height=5cm]{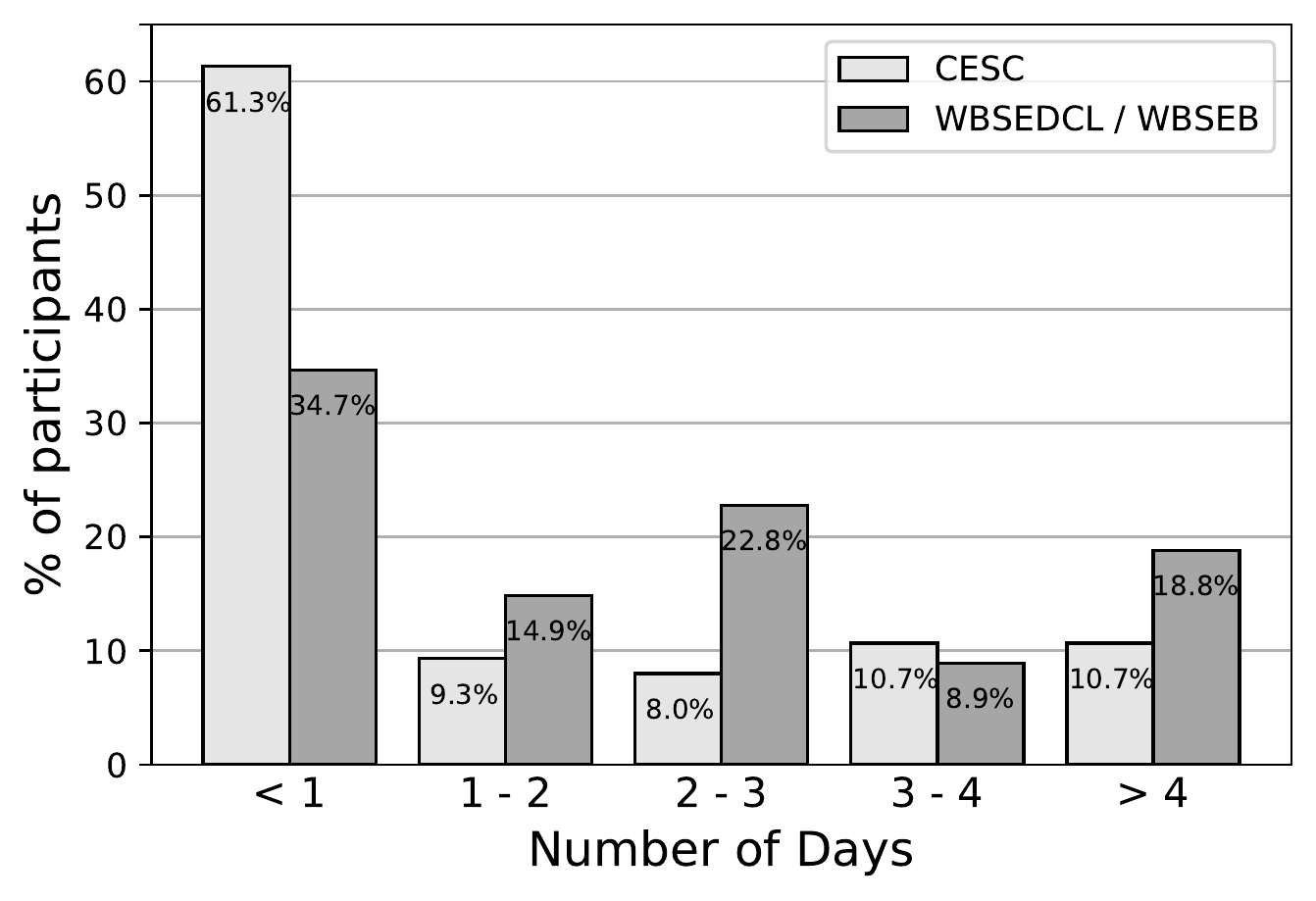}
    \caption{\bf Comparison of electricity restoration time between CESC and WBSEDCL/WBSEB.}
    \label{fig5:cescwbsebrestore}
\end{subfigure}%

\caption{\bf Time taken (in days) for restoration of electricity supply. We present (a)~overall results, and (b)~a comparison between the two major electricity suppliers -- CESC and WBSEB/WBSEDCL. The percentages in the bar charts are calculated based on the number of participants for whom services were disrupted (91 participants supplied by CESC, and 106 by WBSEDCL/WBSEB; 82.4\% and 95.3\% participants respectively faced disruption for them).}
\label{fig5:allelectricrestore}
\end{figure*}

The distribution of time taken for restoration of electricity supply is given in Figure~\ref{fig5:allelectricrestore}. The overall trends are given in Figure~\ref{fig5:electricrestore} whereas Figure~\ref{fig5:cescwbsebrestore} gives a comparison between the two major electricity suppliers CESC and WBSEDCL/WBSEB. 
In case of CESC, the majority of the connections were restored within a day (61\%), whereas WBSEDCL/WBSEB took longer to restore supplies. However for both providers, a considerable fraction of connections took longer than 3 days to restore (21.4\% for CESC, and 27.6\% for WBSEDCL/WBSEB). 

The district-wise distribution of disruption of electric supply is given in Table~\ref{tab3:districtdamage}. As we can see, a significantly less number of participants from Kolkata (25.4\%) faced disruption beyond 1 day. The ``Others'' were not affected too much by the cyclone, hence most of the participants from these regions had electricity restored in under a day. 
However, the districts of North 24 Parganas, South 24 Parganas, Hooghly and Howrah saw heavy disruption of electricity services. 

It can be noted that a large majority of electricity connections of Kolkata and Howrah are supplied by CESC. There is a huge difference in the fraction of connections in these two districts that were disrupted for more than a day -- 25.4\% for Kolkata and 64.3\% for Howrah. This difference indicates that CESC may 
%Many of our participants were from Kolkata where CESC is the majority supplier (96.6\% participants from Kolkata said they were supplied by CESC), and they may
have prioritised fixing connections in Kolkata since it is a major city.

We had also asked the participants (who said they faced disruption of electricity supply) {\it if their locality had some public utility whose operation could have been disrupted due to lack of electricity}. About 11.7\% participants answered there were {\it hospitals/nursing homes} in their localities,  26.1\% said there were {\it water pumping stations}, and as many as 67.8\% participants said there were {\it mobile towers} in their locality. 
The mobile towers not getting electricity supply for long could have depleted their backup power source, which led to further disruption of phone and internet services (as discussed below). The same goes for the water pumping stations, whose disruption could have led to unavailability of drinking water in certain localities for a long duration.

% \begin{table*}[!ht]
%     \centering
%     \normalsize
%     \begin{tabular}{|l|c|c|c|c|c|c|}
%         \hline
%         Supplier & Total & Disrupted & \multicolumn{4}{|c|}{Among the ones disrupted, service was restored in}\\
%         \cline{4-7}
%         & & & \textless 1 day & 1-2 days & 2-3 days & \textgreater 3 days \\
%         \hline
%         CESC & 91 & 82.4\% & 61.3\% & 9.3\% & 8\% & 21.4\% \\
%         WBSEB/WBSEDCL & 106 & 95.3\% & 34.7\% & 14.9\% & 22.8\% & 27.6\% \\
%         \hline
%     \end{tabular}
%     \caption{Electricity Supplier and the restoration time for services}
%     \label{tab5:electricitysupplier}
% \end{table*}

\vspace{2mm}
\noindent \textbf{Disruption of phone and Internet Services:} Next we asked about the disruptions to phone and internet services. For the phone connections, we asked survey participants to mark which service providers they used, and which of them were disrupted, and how long it took for at least one service to be restored. For Internet Connections we asked which media they used to access the internet (like mobile internet or broadband), and if they faced disruption of internet services.

It can be noted that there are four primary mobile service providers in the state of West Bengal -- Vodafone, Airtel, BSNL and Jio. Out of the 201 survey participants, 114, 103, 53, and 132 respectively reported using these services. Note that many participants reported to be using more than one mobile service providers simultaneously. For Internet connectivity, there are two popular choices -- mobile Internet (used by 168 participants), and broadband Internet (used by 111 participants). Again, several participants reported to be using both.

%which are given in Table~\ref{tab5:phonecarrier} and Table~\ref{tab5:internetsupplier} respectively. 

Table~\ref{tab5:phonecarrier} shows the fraction of mobile connections that were disrupted.
Overall, a huge number (85\%) of phone connections seem to have been disrupted, with even more (88.6\%) internet connections being disrupted. Among phone service providers, Vodafone seems to have been disrupted the most (88\%), followed by Airtel (78.6\%), while Jio connections were disrupted quite less relatively (52\%). 
% even though the number of people having Jio connections is the most. 
%\hl{The possible reason for this could've been that Jio is relatively quite new service, which makes their hardware setup and equipment much newer, and thus was able to be disrupted less due to the cyclone}. 
These observations in our survey corroborate with reports in the news media. For instance, The Indian Express reported\footnote{\url{https://tinyurl.com/IE-amphan}} on May 22, 2020 (i.e., two days after the cyclone) -- ``indianexpress.com reached out to some people in Kolkata, and most of them told us that only Jio and BSNL services are working somewhat decently, whereas, Vodafone and Airtel services are not working at all''.

Table~\ref{tab5:internetsupplier} shows the fraction of disrupted Internet connections. 
Internet connections were also heavily disrupted, where close to 90\% of both mobile Internet connections as well as Broadband Internet connections were disrupted. 

\begin{figure}[hbt]
    \centering
    \includegraphics[width =0.9\linewidth,height=5cm]{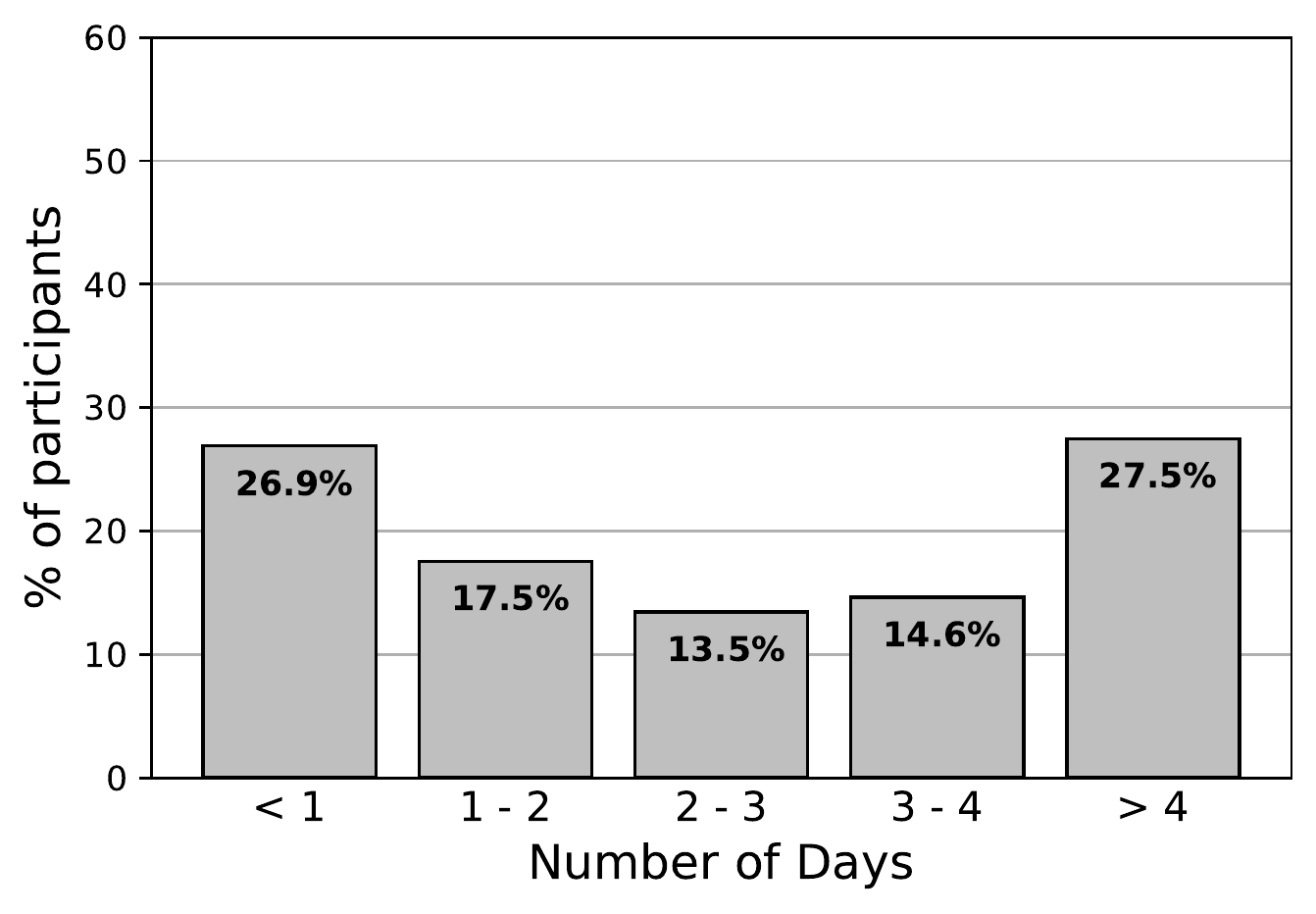}
    \caption{\bf Time taken for phone services to be restored.}
    \label{fig5:phonerestore}
\end{figure}

\begin{figure}[hbt]
    \centering
    \includegraphics[width =0.9\linewidth,height=5cm]{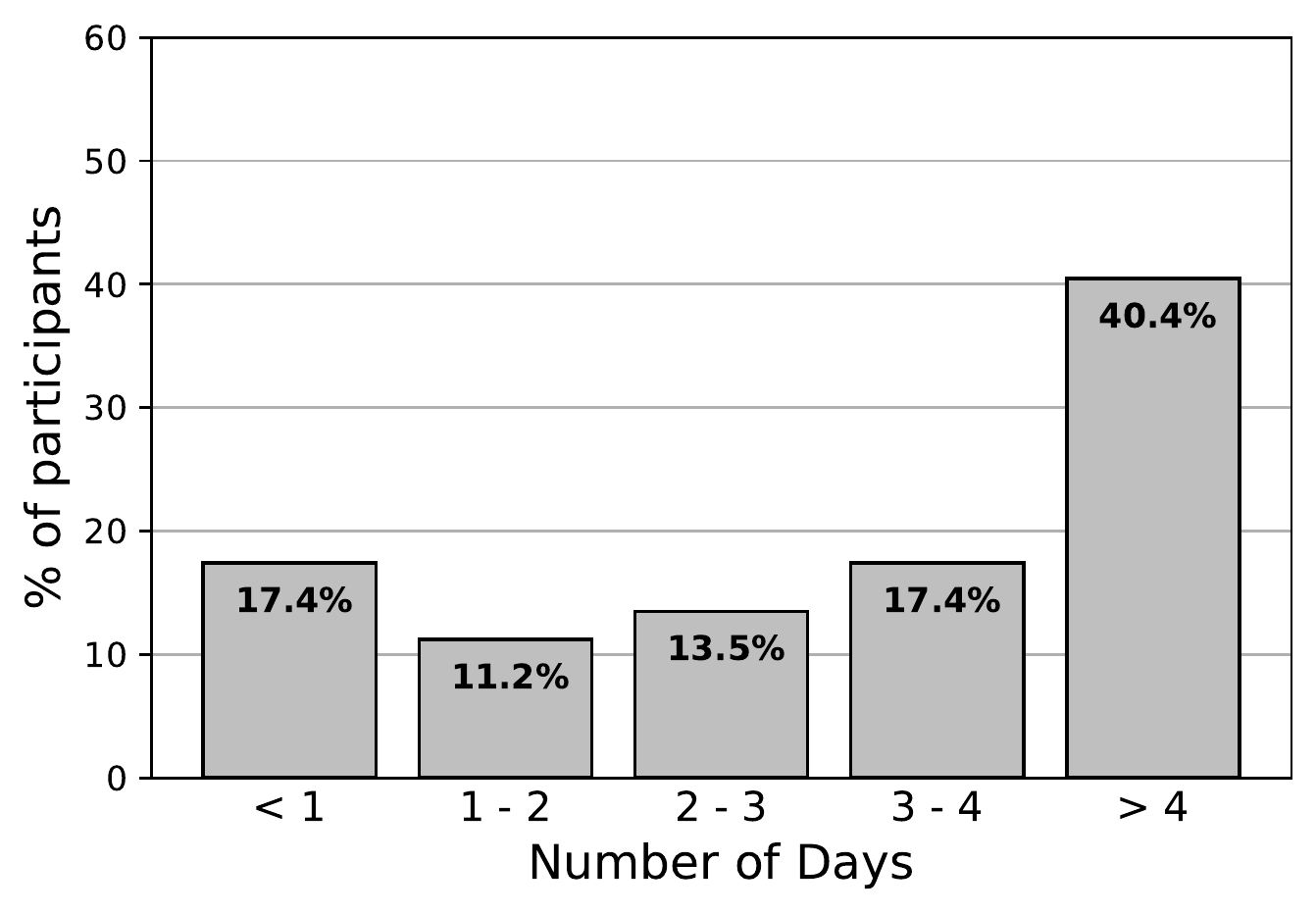}
    \caption{\bf Time taken for internet services to be restored. Note that restoration of internet services took more time than restoration of phone services, for a substantial fraction of participants.}
    \label{fig5:internetrestore}
\end{figure}

The distribution of {\it restoration times} of the phone services is given in Figure~\ref{fig5:phonerestore} where it is seen that a large fraction of phone connections (42.1\%) took more than 3 days to be restored. 
The distribution of restoration time of internet services is given is Figure~\ref{fig5:internetrestore} and a significant fraction of them (57.8\%) took more than 3 days to be restored. 

The district-wise distributions of phone and internet connections that were disrupted for more than a day, are shown in Table~\ref{tab3:districtdamage}. 
It seems the phone and internet services were disrupted for over a day for a large majority of participants in all the five districts that were mostly affected by Amphan. The districts of Kolkata, South 24 Parganas and Howrah (especially for internet service disruption) seem to have been severely affected (though our sample size from Howrah and South 24 Parganas is quite small).

It can be noted that phone connections were disrupted slightly less than internet connections. Several participants indicated that, even though they could not access the internet through their mobile connections, they could use their phones for calling (especially the ones using the service provider Jio).

%\textit{\textbf{Note:} that many of the responders said that they / their families use multiple phone service providers. In cases where a responder used multiple phone service providers, we asked the time of service restoration for the service provider that resumed earliest. E.g., if a responder said he/she uses both BSNL and Jio, and said that the provider that resumed earliest was restored in 12 hours, then we assume both BSNL and Jio were restored within 12 hours for this responder. Hence the estimates given below are optimistic ones for individual service providers.}

\begin{table}[hbt]
    \centering
    \normalsize
    \begin{tabular}{|l|c|c|}
        \hline
        \bf Network Provider & \bf Total & \bf Disrupted \\
        \hline\hline 
        Overall & 201 & 85.1\% \\ 
        \hline \hline
        Vodafone & 114 & 88.6\% \\
        Airtel & 103 & 78.6\% \\
        BSNL & 53 & 66.0\% \\
        Jio & 132 & 52.2\% \\
        \hline
    \end{tabular}
    \caption{\bf Number of participants who avail different network providers for phone and percentage of participants who experienced disrupted phone services.}
    \label{tab5:phonecarrier}
\end{table}

\begin{table}[tb]
    \centering
    \normalsize
    \begin{tabular}{|l|c|c|}
        \hline
        \bf Internet technology & \bf Total & \bf Disrupted \\
        \hline \hline
        Overall & 201 & 88.6\% \\
        \hline \hline
        Mobile & 168 & 89.9\% \\
        Broadband & 111 & 90.1\% \\
        \hline
    \end{tabular}
    \caption{\bf Number of participants who avail internet via different technologies, and percentage of participants who experienced disrupted internet services.}
    \label{tab5:internetsupplier}
\end{table}

\subsection*{Summary of Section}

\noindent The major physical damage (in the localities covered by our survey participants) was uprooting of trees, whereas the damage to buildings seemed less (since most of the participants were from urban areas). 
There was not too much damage in terms of waterlogging and disruption of supply of drinking water, even though some participants faced disruption of drinking water supply beyond 3 days. The district of Hooghly seems to have faced the most damage, however the number of participants from here is quite less. 

Importantly, a huge fraction of participants faced disruption of electricity, phone and internet services; even worse, a huge fraction of these services took more than 3 days to be restored (this delay in restoration could have been higher due to the ongoing COVID19-induced lockdown).
Among the two most popular electricity suppliers, it seems that CESC managed to restore a significantly higher fraction of connections within a day after the cyclone, as compared to WBSEDCL.
Among the popular phone service providers, Jio seems to have faced relatively less disruption.

\section{Effect of these damages on the population}
\label{affected}

\begin{figure*}[hbt]
\begin{subfigure}{0.45\textwidth}
    \centering
    \includegraphics[width =0.9\linewidth]{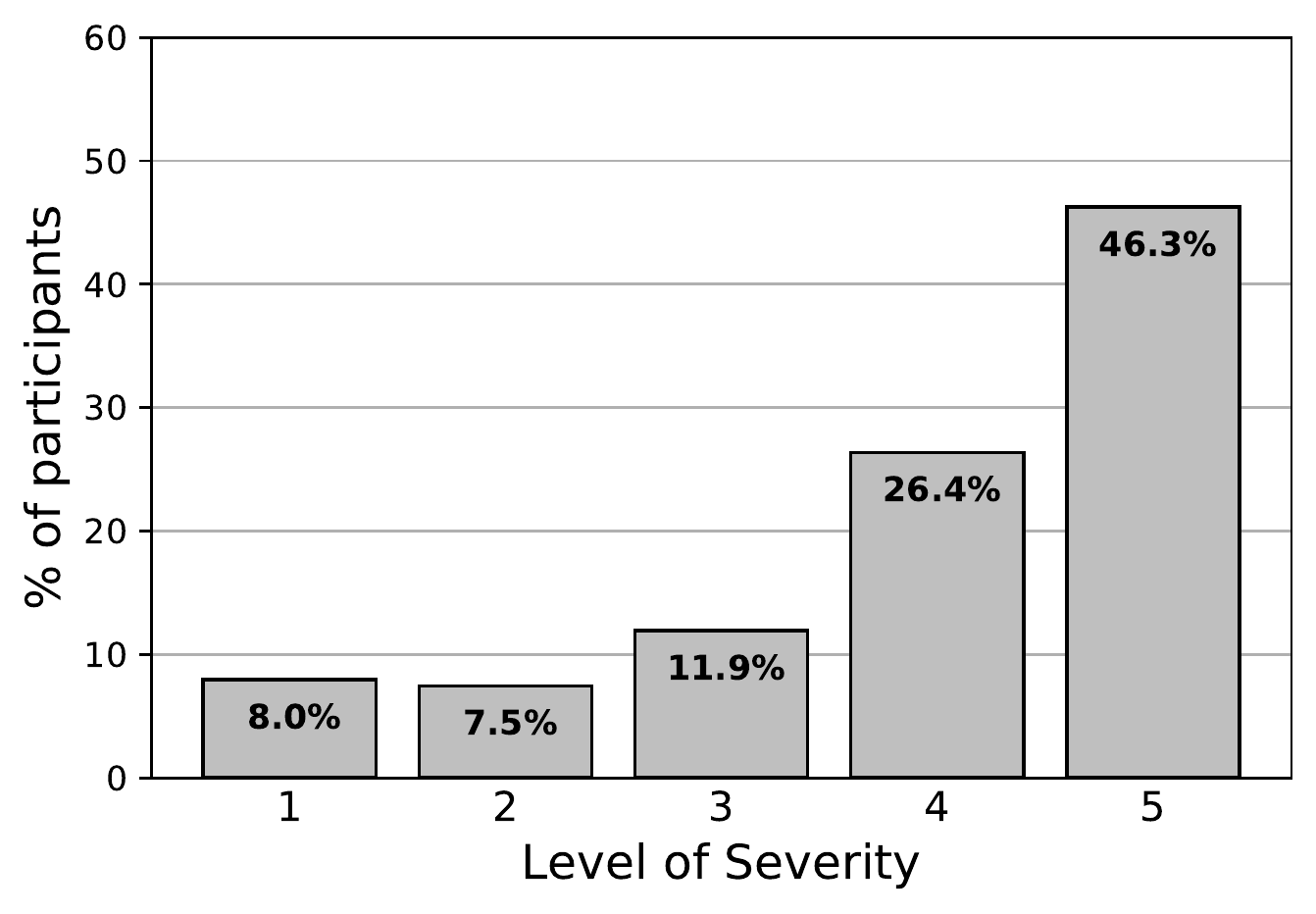}
    \caption{\bf Self-reported severity for disruption of internet services.}
    \label{fig4:internetaffected}
\end{subfigure}
\hfill
\begin{subfigure}{0.45\textwidth}
    \centering
    \includegraphics[width =0.9\linewidth,height=5cm]{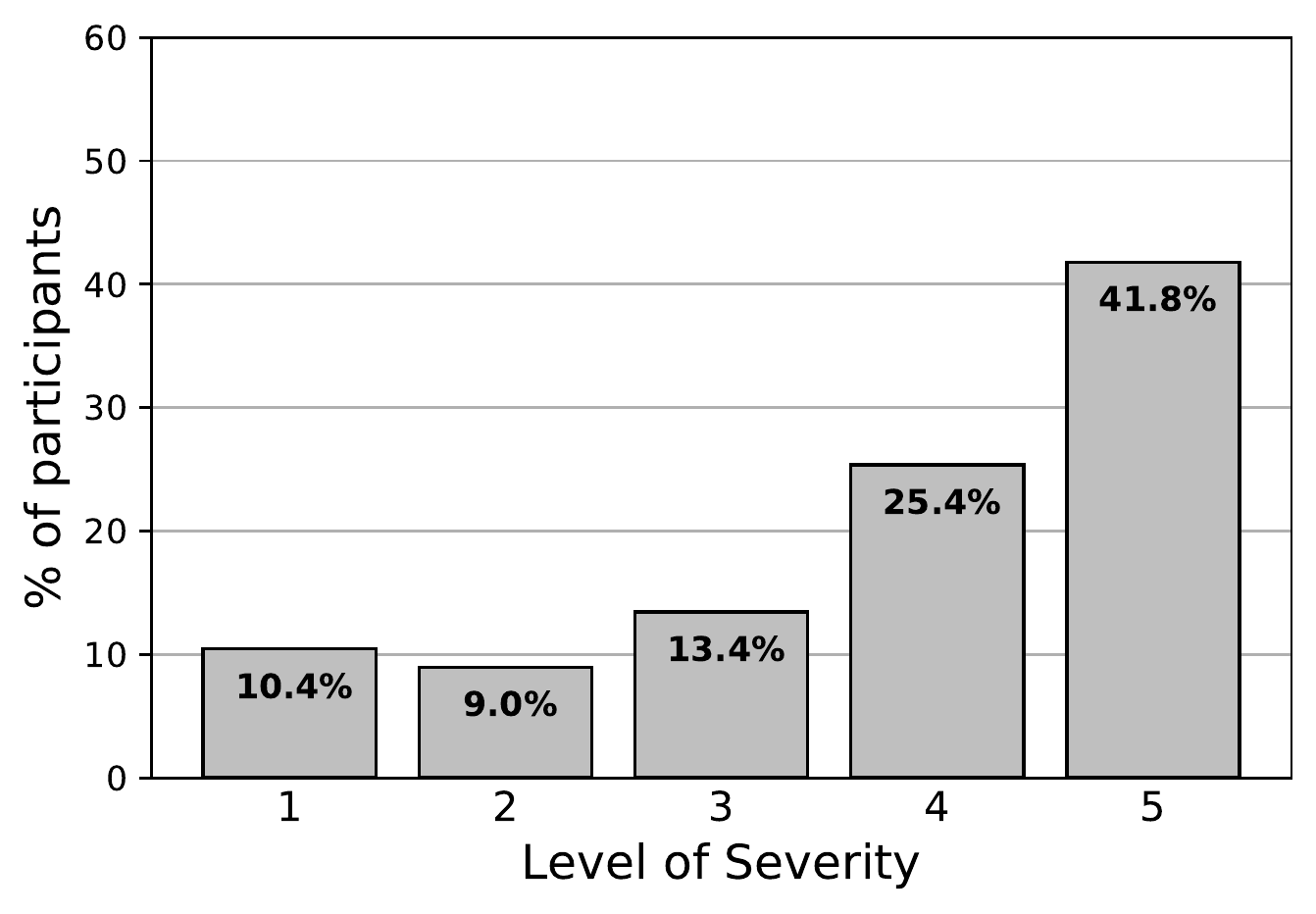}
    \caption{\bf Self-reported severity for disruption of phone services.}
    \label{fig4:phoneaffected}
\end{subfigure} % 
\newline
\begin{subfigure}{0.45\textwidth}
    \centering
    \includegraphics[width =0.9\linewidth,height=5cm]{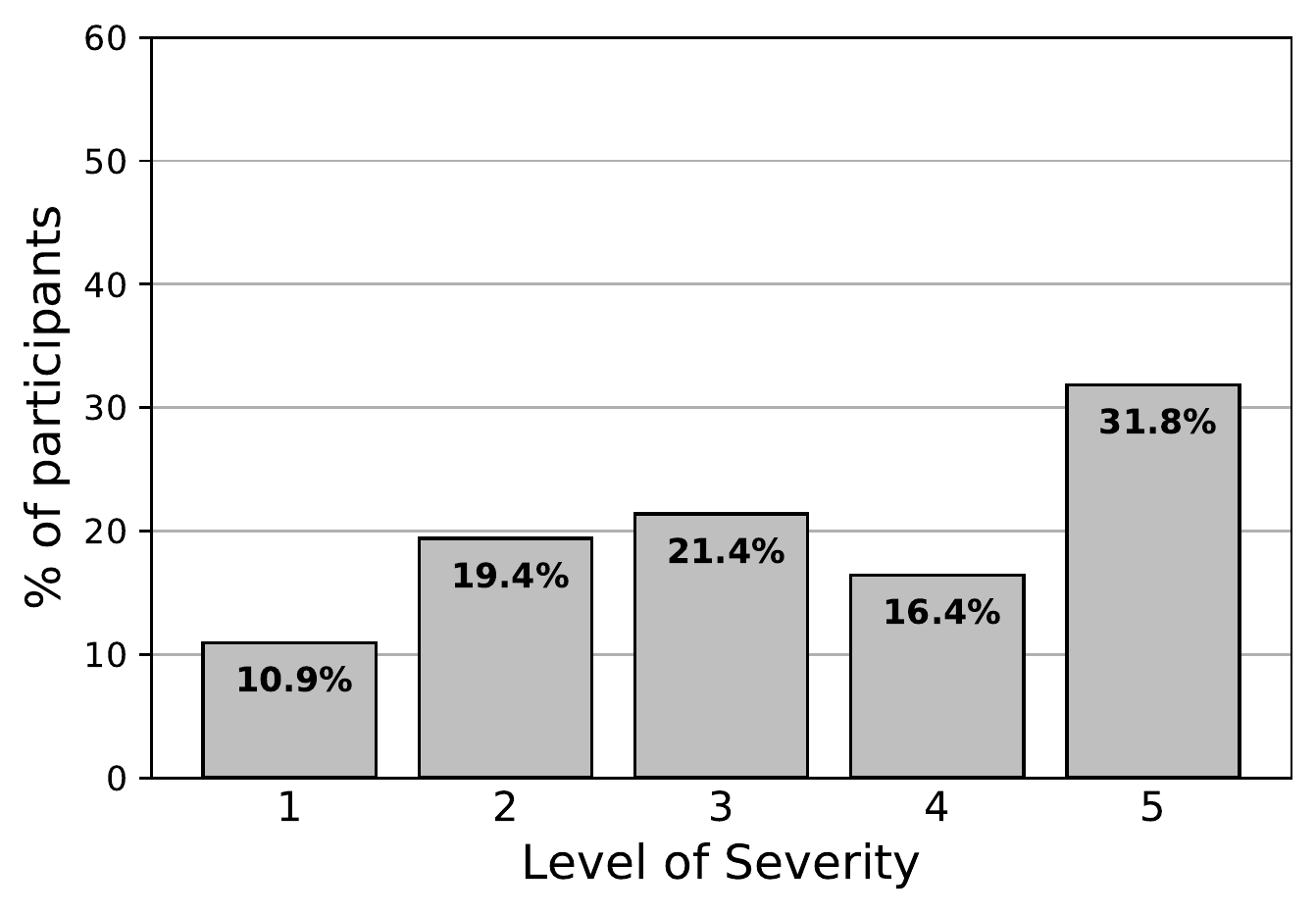}
    \caption{\bf Self-reported severity for disruption of electricity supply.}
    \label{fig4:electricaffected}
\end{subfigure}
\hfill
\begin{subfigure}{0.45\textwidth}
    \centering
    \includegraphics[width =0.9\linewidth,height=5cm]{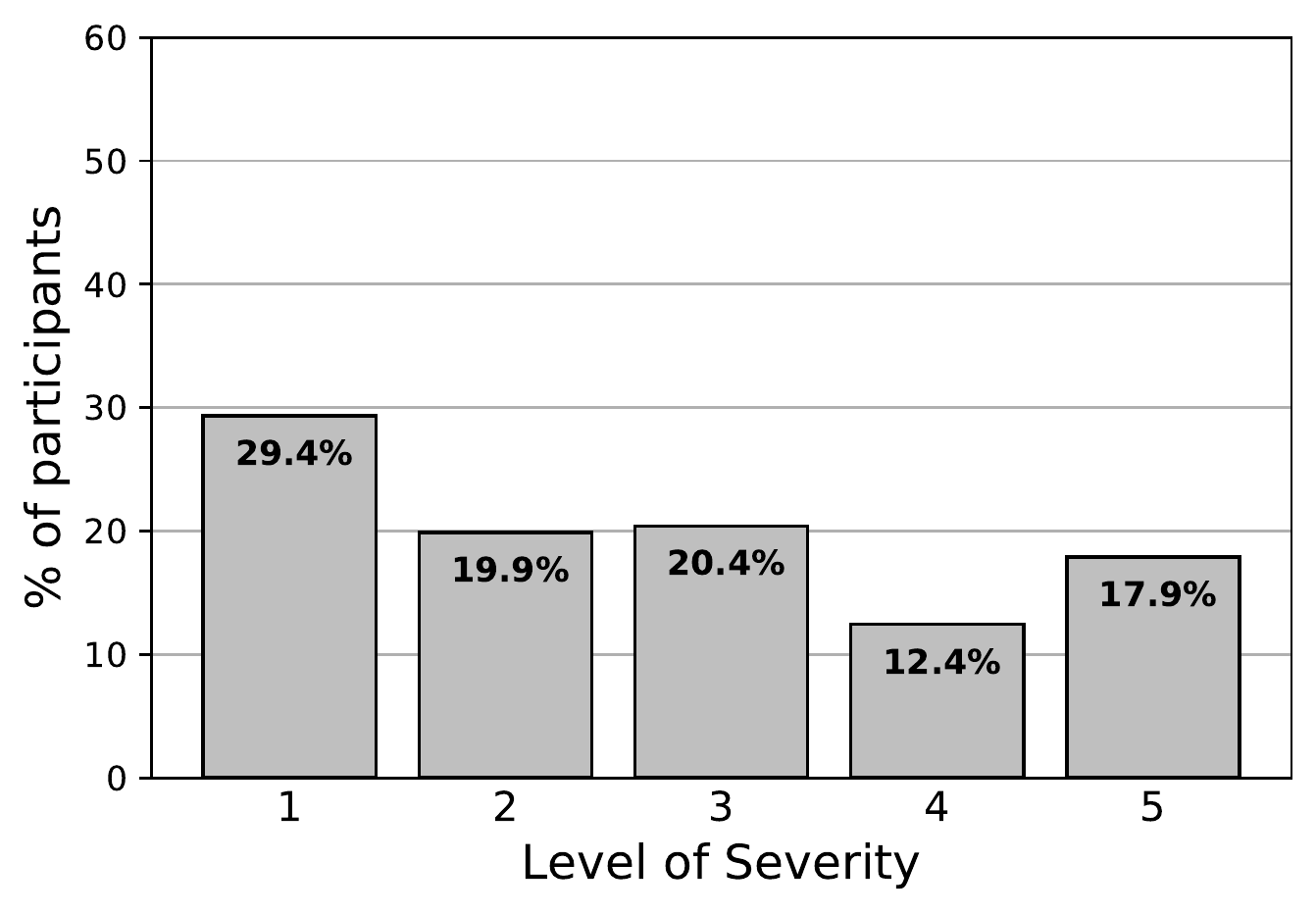}
    \caption{\bf Self-reported severity for uprooting of trees.}
    \label{fig4:treesaffected}
\end{subfigure}
\newline
\begin{subfigure}{0.32\textwidth}
    \centering
    \includegraphics[width =\linewidth]{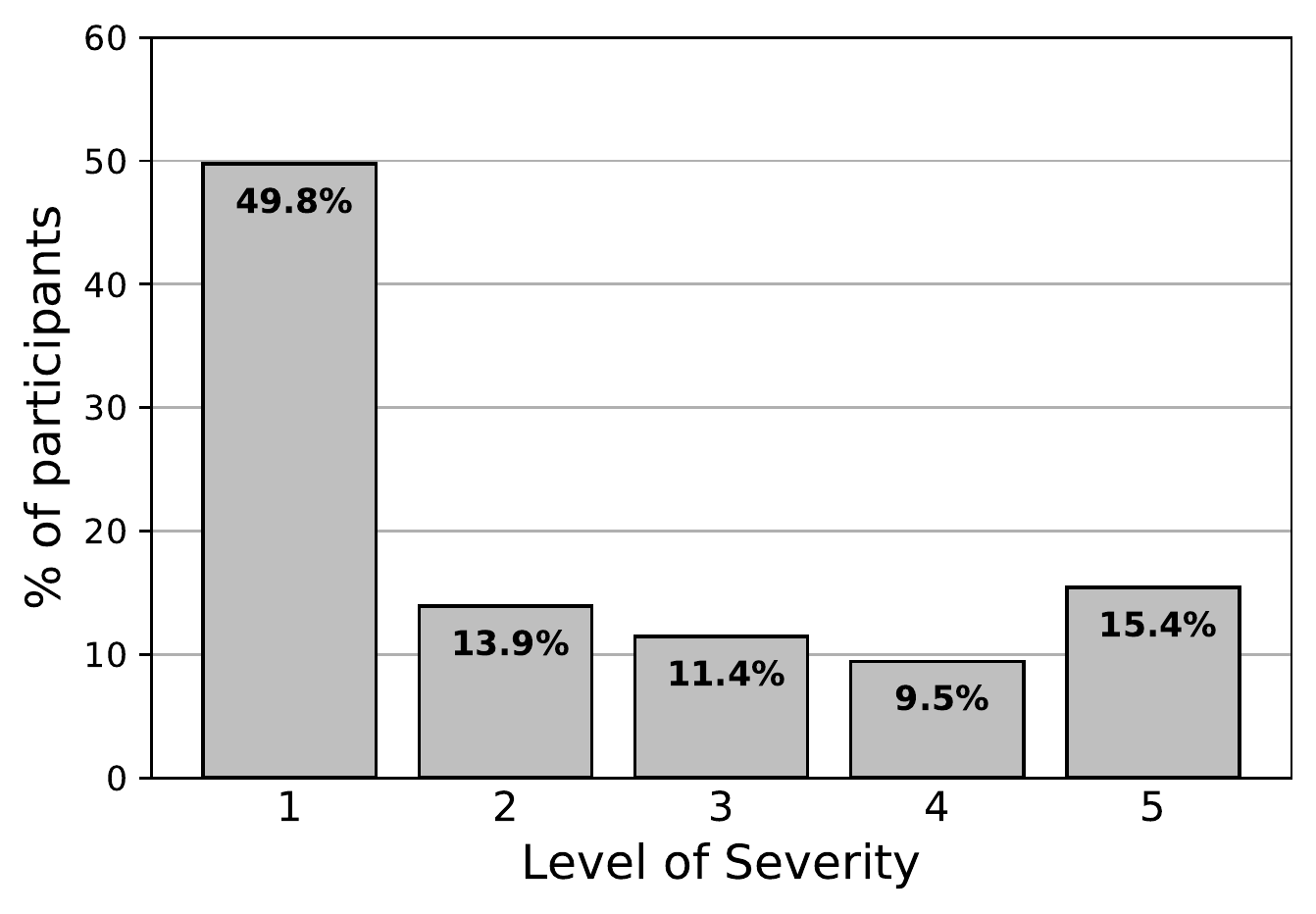}
    \caption{\bf Self-reported severity for disruption of supply of drinking water.}
    \label{fig4:drinccaffected}
\end{subfigure}%
\hfill
\begin{subfigure}{0.32\textwidth}
    \centering
    \includegraphics[width =\linewidth]{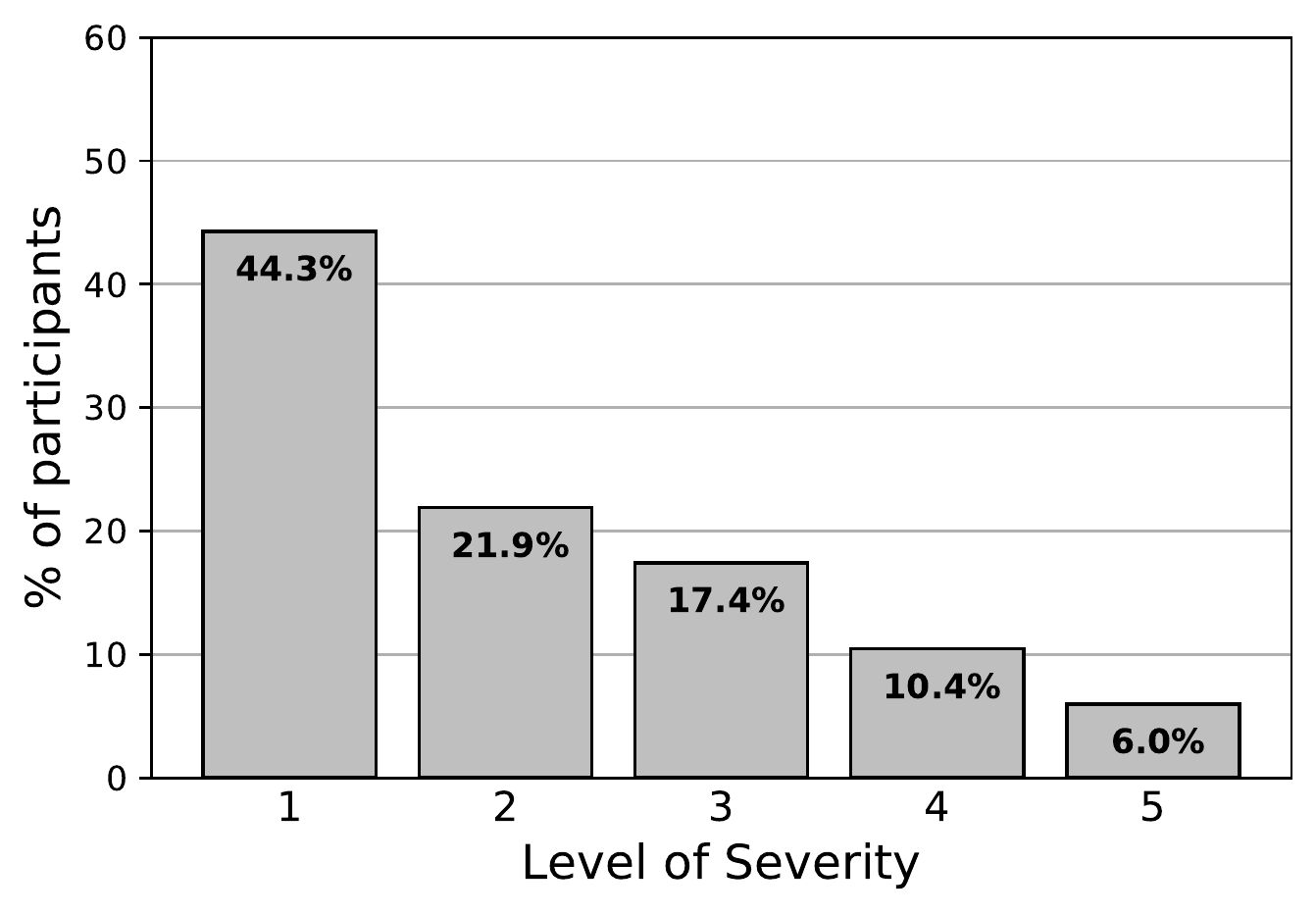}
    \caption{\bf Self-reported severity for to waterlogging.}
    \label{fig4:wateraffected}
\end{subfigure}%
\hfill
\begin{subfigure}{0.32\textwidth}
    \centering
    \includegraphics[width =\linewidth]{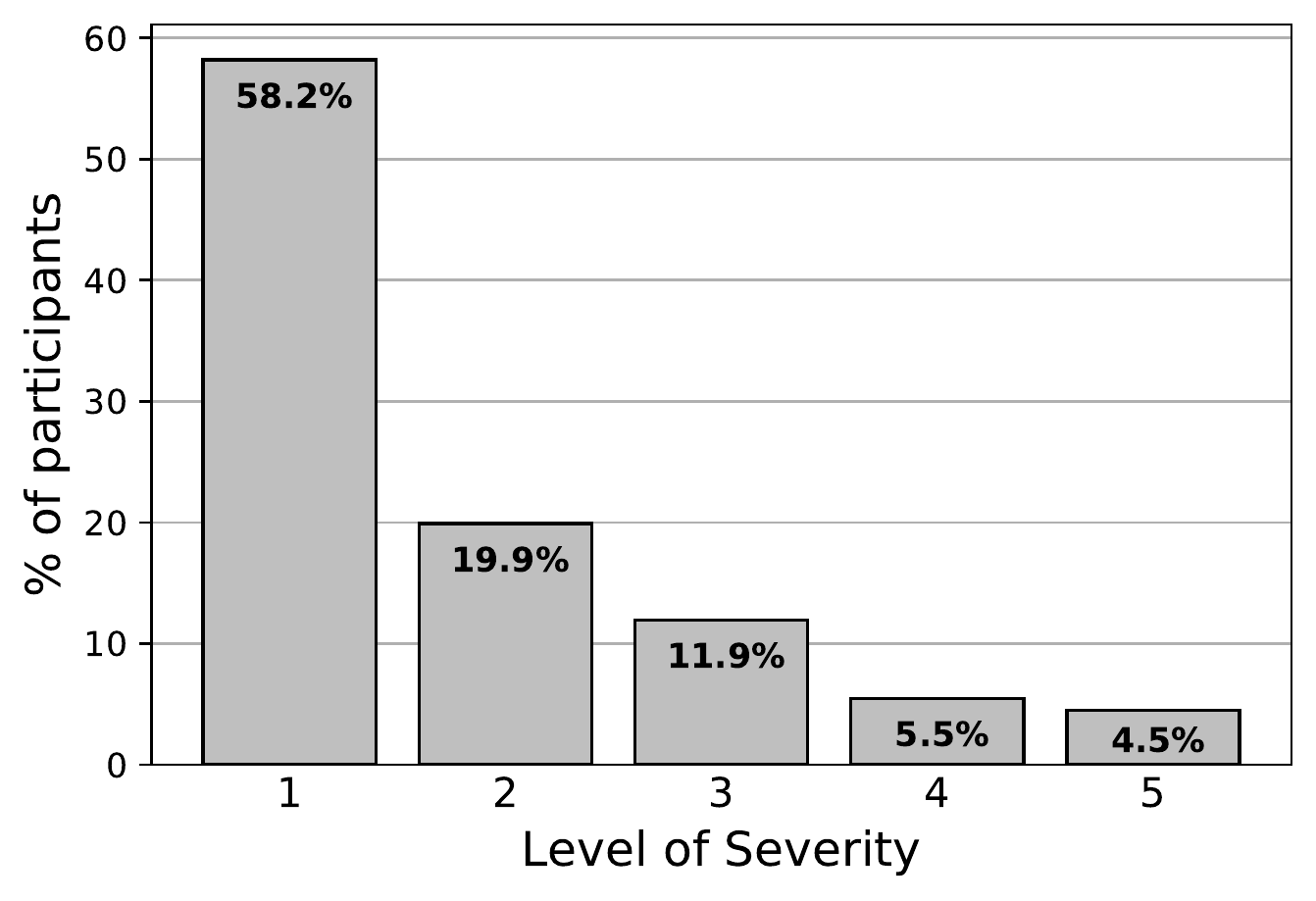}
    \caption{\bf Self-reported severity for damages to buildings.}
    \label{fig4:buildingaffected}
\end{subfigure}%
\caption{\bf Severity of participants affected by disruption of various factors on a scale of 1 (least severe) to 5 (most severe). Participants were affected much more by disruption of services than physical damages to their localities.}
\label{fig4:allaffected}
\end{figure*}

\begin{table*}[hbt]
    \centering
    \normalsize
    \begin{tabular}{|l||c||c|c|c|c|c|c|}
        \hline
        \shortstack[l]{\bf Participants who said they were\\ \bf heavily affected (levels 4-5) due to} & 
        \shortstack{\bf Overall\\(201)} & \shortstack{\bf N24P\\(60)} & \shortstack{\bf KOL\\(59)} & \shortstack{\bf S24P\\(22)} & \shortstack{\bf HGLY\\(18)} & \shortstack{\bf HWH\\(14)} & \shortstack{\bf Others\\(28)} \\
        \hline \hline 
        Disruption of Internet Connectivity & 72.8\% &  81.7\% & 69.5\% & 86.4\% & 77.8\% & \textbf{100\%} & 32.1\%\\
        
        Disruption of Phone Connectivity & 67.3\% & 73.3\% & 67.8\% & \textbf{86.4\%} & 72.2\% & 78.6\% & 28.6\% \\
        
        Disruption of Electricity Supply & 48.5\% & 60.6\% & 27.1\% & 77.3\% & 50.0\% & \textbf{78.6\%} & 28.6\% \\
        
        %\hline

        Uprooting of Trees & 30.0\% & 33.3\% & 23.7\% & \textbf{50.0\%} & 33.3\% & 35.7\% & 17.9\% \\

        Disruption of Drinking Water Supply & 24.7\% & 30.0\% & 20.3\% & 31.8\% & 22.2\% & \textbf{42.9\%} & 10.7\% \\        

        Waterlogging / Flooding & 16.3\% & 20.0\% & \textbf{22.0\%} & 9.1\% & 11.1\% & 21.4\% & 3.6\% \\

        Damage to Buildings & 9.9\% & \textbf{16.7\%} & 6.8\% & 9.1\% & 5.6\% & 7.1\% & 7.1\% \\
        \hline
    \end{tabular}
    \caption{\bf Distribution of participants who were severely affected (i.e., selected one of the highest two levels 4 or 5 on Likert scale of being affected) by various factors, across the different districts. The percentages represent the fraction of participants who were severely affected from that district. The numbers in brackets below each of the headers represent the count of participants from that district. The highest percentage in each row is highlighted in boldface. The seven factors (rows) are arranged in decreasing order of the overall fraction of participants who said they were severely affected by a certain factor ($2^{nd}$ column).}
    \label{tab4:districtaffected}
\end{table*}

\noindent In this section, we attempt to study how severely the population was affected by the damages caused to their localities. We considered the seven factors discussed in Section~\ref{damages}, and tried to find to what extent the participants were affected by each factor. %We also try to find out which of the factors affect participants more, and need to be focused on by the authorities to mitigate the damages in future scenarios. 

To this end, for each of the seven factors, we asked participants to rate the extent to which they / their family were affected, on a {\it Likert scale} of 1-5, where level `1' indicates `not affected at all', and level `5' indicates `extremely severely affected'. 
The distributions of the responses for all seven factors are shown in Figure~\ref{fig4:allaffected}(a)-(g).

We consider a participant to be \textit{``severely affected''} if he/she selected levels 4 or 5 on the Likert scale. We have sorted the factors based on the decreasing order of number of people saying that they were severely affected by a certain factor. 
We also calculated which factors affected people severely (participants who selected levels 4-5) from each of the districts, and have tabulated the district-wise distributions in Table~\ref{tab4:districtaffected}.

\vspace{2mm}
\noindent \textbf{Electricity, phone, Internet and drinking water services:} 
A staggering 72.8\% of the participants were severely affected (i.e., selected levels 4 or 5 on the Likert scale) by disruption of internet services (see Figure~\ref{fig4:internetaffected}).
As many as 67.3\% of the participants were severely affected by disruption in phone services (Figure~\ref{fig4:phoneaffected}) while 48.5\% of the participants were severely affected by disruption of electricity supply (Figure~\ref{fig4:electricaffected}).

There can be two potential factors leading to such high fractions of people getting severely affected by disruptions of these services. First, as stated earlier, the cyclone occurred during the COVID19-induced lockdown which forced a lot of people to work from home. 
Second, our sample population mostly consisted of students, faculty and engineers (as given in Section~\ref{demographics}). These services were particularly crucial for our sample population while working from home. 
These two factors might explain why the disruption severely affected so many of the survey participants.

The district-wise distribution of these participants is given in the first three rows of Table~\ref{tab4:districtaffected}. It seems that people from Howrah and South 24 Parganas were most severely affected overall (though the number of participants from these district are only 14 and 22 respectively). 
%Most number of participants facing problems due to internet (49) and phone service (44) disruption are from North 24 Paraganas whereas percentage-wise, South 24 Paraganas and Howrah are affected the most. Similar trends can be seen for disruption of electricity supply.

Also 24.7\% of the participants were severely affected by drinking water shortage (Figure~\ref{fig3:drinccdamage}), with the highest fraction being from Howrah.

\vspace{2mm}
\noindent \textbf{Physical damages to localities:} 
For the three physical damage factors (uprooting of trees, damage to buildings, and waterlogging), the number of people severely affected follows the same pattern as the amount of damages caused. 
Figure~\ref{fig4:treesaffected} shows that 30.3\% of the participants were severely affected due to uprooting of trees. 
Figure~\ref{fig4:wateraffected} shows that not too many (16.3\%) were affected severely by waterlogging and even fewer (10\%) by damages to buildings as shown in Figure~\ref{fig4:buildingaffected}. 
The probable reason for these low fractions is that the participants were from urban areas mostly. 

\if 0
\red{I suggest to remove this paragraph. See if you agree -- SG}
%Note that, in Section~\ref{damages} we saw that a lot of trees had been uprooted, but it seems people were not affected much by this factor. The possible reason again comes down to the COVID19-induced lockdown, when people were not moving out of their homes, and thus were not much affected by uprooted tress (unless the uprooted trees disrupted the other services, for example if the uprooted trees had fallen on power lines causing disruption of electricity connection). 
%A similar reason can be given for people not getting affected too severely by waterlogging.
\fi 

The district-wise distribution of these participants is given in Table~\ref{tab4:districtaffected}. Uprooting of trees seems to have affected people more or less equally from all the districts, except for South 24 paraganas where a lot more people (50\%) seem to have been affected, which is expected because of the huge amount of trees present in this district. Waterlogging seems to have been a problem mostly in Kolkata, North 24 Paraganas, and Howrah. 
North 24 Paraganas seems to be most affected by damage to buildings, but still only a few participants ($10$) said they were affected severely.

% \hl{DO WE KEEP THIS?}\\
% Electricity Supplier for Howrah: 50-50 CESC and WBSEB\\
% Electricity Supplier for Kolkata: 96.6\% CESC\\
% Electricity Supplier for N24Pgs: 20\% CESC, 78.3\% WBSEB  (others: 1.7\%)\\

% \subsection*{Correlation of Heavily affected with Damage to Locality}
% \textbf{Trees Uprooted:}  Support: 61 participants said heavily affected due to uprooting of trees. Out of these 61, 71\% said more than 10 trees were uprooted in their locality and 15\% said between 6 to 10 trees were uprooted in their locality. \\
% More than 10 trees: 77\% \\
% Between 6 to 10: 14.8\% \\
% Rest: 8.2\% \\

% \textbf{Buildings Damages:} Support: 20 \\
% More than 10 buildings: 45\% \\
% Between 6 to 10: 15\% \\
% Between 1 to 5: 30\% \\
% Rest: 10\% \\

% \textbf{Waterlogging:} Support: 33 \\
% More than 2 days: 30.3\% \\
% Between 1 and 2 days: 24.2\% \\
% Between 12 hours to 1 day: 15.2 \% \\
% Rest: 30.3\% \\

% \textbf{Drinking Water Shortage:} Support: 50 participants said heavily affected due to drinking water shortage. Out of these 50, 52\% said drinking water supply was not regular for more than 2 days, 26\% said drinking water supply was not regular between 1 and 2 days. \\
% More than 2 days: 52\% \\
% Between 1 and 2 days: 26\%	\\
% Between 12 hours to 1 day: 16\% \\
% Rest: 6\% \\

\subsection*{Summary of Section}
\noindent The population was affected much more by disruption of electricity, phone and internet services than by physical damages such as uprooting of trees and waterlogging 
(which may be because of the COVID19-induced lockdown and the bias in our survey population). 
All districts seem to have been more or less equally affected due to disruption of internet and phone services (except for ``Others'' which were anyway affected less by the cyclone). 
South 24 Paraganas and Howrah seem to have been affected more by disruption of electricity supply. Participants from South 24 Paraganas seem to have been affected significantly more due to uprooting of trees, whereas Kolkata and North 24 Paraganas seem to have been affected more due to waterlogging.

    % \item For correlation between people affected heavily(4-5) and damages to the locality, it seems mostly it follows the expected pattern. Two questions arise:
    % \begin{itemize}
    %     \item In Buildings damaged, quite a few people (30\%) who have stated they have been severely affected say that there are less than 5 buildings damaged in their locality. This discrepancy may be because the houses of these participants were specifically damaged.
    %     \item For all the cases, there are also few people (about 8\%, and 30\% in waterlogging) who have said that they were heavily affected but have stated there was no damage to the locality (or “cannot say”). Why? [Possible inconsistency]
    % \end{itemize}

\section{Impact of Social Media}
\label{socialmedia}

\noindent 
Online Social Media (OSM) is being increasingly used in post-disaster times both within India and abroad~\cite{10.1145/3137597.3137602,10.1145/2771588,osm-usage-disaster,osm-emergencies-15yrs}.
In this section, we analyse the impact of OSM on people during the days immediately after the Amphan cyclone. 
%We study the posts that have been posted by our participants and the ones that they have seen on social media. 
We also try to analyse how effective the authorities were in utilising social media to reach people during the disaster.

\begin{figure}[hbt]
    \centering
    \includegraphics[width =0.9\linewidth,height=5.5cm]{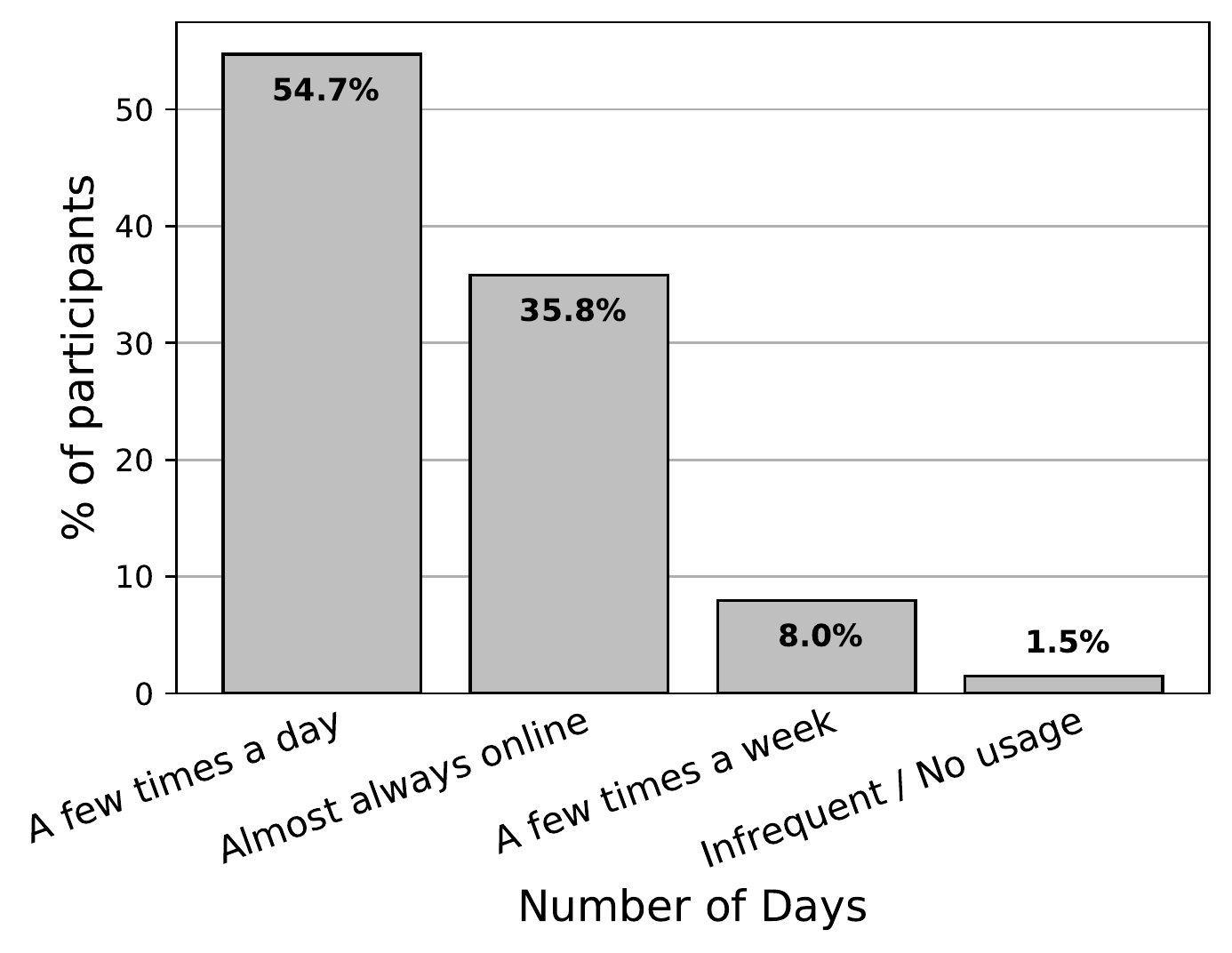}
    \caption{\bf Frequency of social media usage of our participants.}
    \label{fig6:socialusage}
\end{figure}

\vspace{2mm}
\noindent \textbf{OSM usage of participants in general:} Out of the total of 201 survey participants, 198 said that they use social media, and the frequency of usage is as given in Fig.~\ref{fig6:socialusage}. Our survey participants mostly consist of people who frequently use social media, with 90.5\% participants using social media every day. 
This high social media usage is expected as our survey population is biased towards people who frequently use social media and are thus more likely to respond to our survey (as discussed in Section~\ref{demographics}). 

We also asked the participants which social media apps/sites they use, with a lot of them using Facebook and Whatsapp as shown in in Table~\ref{tab6:popularsocial}.

\begin{table}[!ht]
    \centering
    \normalsize
    \begin{tabular}{|c|c|c|c|}
        \hline
        \bf Twitter & \bf Instagram & \bf Facebook & \bf Whatsapp\\
        \hline
        22.9\% & 46.8\% & 85.1\% & 95.5\% \\
        \hline
    \end{tabular}
    \caption{\bf Different social media used by participants.}
    \label{tab6:popularsocial}
\end{table}

%%%%% table below commented out %%%%
\if 0 

\begin{table}[hbt]
    \centering
    \normalsize
    \begin{tabular}{|c|c|c|}
    \hline
    \multicolumn{3}{|l|}{Frequency of Social Media Usage}\\
    \hline
    Almost Always & Few times a day & Rest \\
    40.9\% & 40.9\% & 18.2\% \\
    \hline
    \multicolumn{3}{|l|}{Time for Internet Connection to be restored}\\
    \hline
    \textless 1 day & Between 1-2 days & \textgreater 2 days \\
    18.2\% & 9.1\% & 72.7\% \\
    \hline
    \multicolumn{3}{|l|}{Time for Electricity Supply to be restored}\\
    \hline
    \textless 1 day & Between 1-2 days & \textgreater 2 days \\
    27.3\% & 9.1\% & 63.6\% \\
    \hline
    \end{tabular}
    \caption{Reasons for the 22 participants not using social media for 7 days}
    \label{tab6:nousage}
\end{table}

\fi 
%%%%%%%% table above commented out %%%%%%%

\begin{table*}[hbt]
    \centering
    \normalsize
    \begin{tabular}{|l|c|c|c|}
        \hline
        \bf Experience with posting information on OSM by  participants & \bf \# & \bf \%  & \bf \% active \\
        \hline\hline
        Connected to Government accounts / NGOs / service providers and asked for help & 10 & 5.0\% &  5.6\% \\
        Posted my opinion on issues related to Amphan & 24 &  11.9\% &  13.4\% \\
        Generally described the situation in my locality & 25 &  12.4\% &  14.0\% \\
        Posted information that can help others (e.g., helpline numbers) & 28 & 13.9\% &  15.6\% \\
        Posted images of the damages in my locality & 36 &  17.9\% &  20.1\% \\ 
        Informed others about safety of myself / my family & 48 &  23.9\% &  26.8\% \\
        Inquired about safety of others & 50 &  24.9\% &  27.9\% \\
        I used social media, but did not post any information related to Amphan & 83 & 41.3\% &  46.4\% \\
        \hline
    \end{tabular} 
    \caption{\bf Experience of posting information on online social media (OSM), as reported by the survey participants. ``\#'' (\%) represent the number (percentage) of participants. ``\% active'' is the fraction of participants who said they were active on social media within the 7 days following Amphan. The rows are ordered in increasing number of participants.}
    \label{tab6:socialposts}
\end{table*}

\begin{table*}[hbt]
    \centering
    \normalsize
    \begin{tabular}{|l|c|c|c|}
        \hline
        \bf Experience with information received by our participants from OSM & \bf \# & \bf  \% & \bf  \% active\\
        \hline \hline
        I used social media, but did not receive any useful information related to Amphan & 33 &  16.4\% &  18.4\% \\
        Received important updates from my own locality & 56 & 27.9\% &  31.3\% \\
        Received useful updates from Government accounts / NGOs / service providers & 59 & 29.4\% &  33.0\% \\
        Received information about safety of others & 77 & 38.3\% &  43.0\%  \\
        Received important updates from other regions & 88 &  43.8\% &  49.2\% \\
        Received advance warning about the Cyclone (from social media) & 126 & 62.7\% &  70.4\% \\
        \hline
    \end{tabular} 
    \caption{\bf Information received by the participants from social media. The columns are same as that of Table~\ref{tab6:socialposts}. The rows are ordered in increasing number of participants.}
    \label{tab6:socialinfo}
\end{table*}

\begin{table*}[hbt]
    \centering
    \normalsize
    \begin{tabular}{|l|c|c|c|}
        \hline
        \bf Experience of undesirable / harmful information on OSM & \bf \# & \bf \%  & \bf \% active\\
        \hline \hline
        Observed too much of religious / superstitious posts & 14 & 7.0\% &  7.8\%  \\
        Observed many posts about other regions, but not enough coverage of my locality & 30 &  14.9\% &  16.8\% \\
        Observed fake news / rumours & 35 & 17.4\% &  19.6\%  \\
        Observed too much useless posts & 38 &  18.9\% &  21.2\% \\
        Observed too much of political arguments & 61 & 30.3\% &  34.1\%  \\
        I used social media, but did not observe any negative effect of social media & 86 & 42.8\% &  48.0\% \\
        \hline
    \end{tabular} 
    \caption{\bf Undesirable / harmful content observed by participants on social media. The columns are same as that of Table~\ref{tab6:socialposts}. The rows are ordered in increasing number of participants.}
    \label{tab6:socialnegative}
\end{table*}

\begin{table*}[hbt]
\centering
\begin{tabular}{|p{0.15\textwidth}|p{0.8\textwidth}|}
    \hline
    \textbf{Category} & \textbf{Responses} \\
    \hline \hline
    \multirow{2}{=}{Situational information} & - Updates on the speed and movement of the cyclone \\
                            & - Updates on the measures taken by the State and Central Government \\
    \hline \hline
    \multirow{4}{=}{Useful contacts} & - Essential phone numbers for connecting CESC, Water Department, Govt. Of WB Facebook Page, etc.\\
                    & - Instagram and WhatsApp stories with helpline nos and CM relief fund donation details\\
                    & - I came across phone numbers where I could address my problems regarding the disruption of electricity, and NGOs helping the needy.\\
    \hline \hline
    \multirow{3}{=}{Attempts to collect donations and relief} & - I could see many relief activities going on in social media and people posting activity pictures and asking for donation to help more. This is a very positive thing I have seen. \\
                    & - Social media prompted to help bring in donation drives from people much faster than other media in these conditions.\\
    \hline \hline
    \multirow{4}{=}{Inquiring about safety of others/self} & -  Social media gave more important updates since TV was not working, and even with less internet connectivity, at least text communication was to some extent helped to assure about relatives.\\
                    & - On Facebook there appeared a button after the cyclone to mark whether I was safe during the cyclone which helps my friends to know about my situation after the cyclone\\
    \hline \hline
    \multirow{3}{=}{Other benefits of social media} & - the hashtags (\#PrayForBengal, \#FightBackbengal etc) expressing togetherness of Bengal were influencing enough.\\
                    & - I feel that the posts, stating the devastating condition of worst-hit area Sundarban due to Amphan are helpful in reality. Many can contribute to the funds which are shared through social media (generally Facebook).\\
    \hline
\end{tabular}
\caption{\bf Sample responses (excerpts) by participants when asked to describe ``social media posts that you found to contain important information on Amphan''.}
\label{tab:exampleposts}
\end{table*}

\begin{table*}[hbt]
\centering
\begin{tabular}{|p{0.8\textwidth}|}
    \hline
    \textbf{Responses} \\
    \hline \hline
    - Posts vilifying little efforts of people trying to donate from selling artwork, trolling unnecessarily for difference of opinions, fake news about politics around relief work etc. But overall negative effect was less than positive.\\
- Fake news and propaganda by political parties\\
- The posts that tried to add a political angle even to a natural disaster like Amphan. \\
- Negativity spreading regarding Electricity disruption\\
    \hline
\end{tabular}
\caption{\bf Sample responses (excerpts) by participants when asked to describe ``social media posts that you think hindered / worsened the situation''.}
\label{tab:examplenegative}
\end{table*}

\vspace{2mm}
\noindent \textbf{OSM usage of participants within 7 days after Amphan:}
We asked three primary questions to the survey participants to find out the impact of social media during and after the disaster -- (i)~whether they posted any information related to cyclone Amphan on OSM, (ii)~whether they received any useful information related to Amphan from OSM, and 
(iii)~whether they observed any undesirable/harmful information related to Amphan on OSM.
All these questions specified that the participants should consider a time window of {\it 7 days after the cyclone} while answering these questions (e.g., whether they had posted/received any information relevant to Amphan from OSM within 7 days after the cyclone).

In each of the 3 questions described previously, we had also kept an option stating ``I did not use social media within 7 days after Amphan''. 
%Surprisingly, the number of participants selecting this option for the three questions was different, ranging between 27 and 32. 22 participants consistently selected this option for all the 3 questions, whereas the others possibly may have marked it wrongly instead of selecting the option that say they used social media but did not engage in posts related to Amphan. 
A total of 22 participants selected this option for all three questions. We tried to find the reasons for these people not using social media. Two potential reasons could be (i)~they are generally infrequent users of OSM, or (ii)~their internet / electricity services were disrupted after the cyclone, thus preventing them from using OSM. 
%The results are stated in~Table \ref{tab6:nousage}.
%22 people said they did not use social media for 7 days (consistently for the 3 questions). 
It is interesting to see that, out of the 22 participants who did not use OSM in 7 days after Amphan, most of them (81.8\%) are frequent users of OSM. 
Among them, about 72.7\% faced internet connection problems and 63.6\% faced electricity problems beyond 2 days after the cyclone, which might have been the reason for them not being able to use social media. 
It is difficult to guess why the rest did not use OSM for those few days. It is possible that they were recuperating from the effects of the cyclone and/or the COVID pandemic.

The rest 179 (89\%) of our survey participants used OSM during the period of 7 days after Amphan. We refer to these participants as `active users'. 
In the rest of this section, we analyze the information posted/received by these active users on OSM over the 7 days immediately after cyclone Amphan.

\vspace{2mm}
\noindent \textbf{Information posted on social media:} 
We first asked {\it if the participants themselves had posted information related to Amphan on social media} during the 7 days immediately after the cyclone. The responses have been tabulated in Table~\ref{tab6:socialposts}. Quite a lot of the participants (46.4\% of the active users) used social media but did not post anything related to Amphan. 
However, the rest actually posted various information related to Amphan on OSM, e.g., described the situation in their locality (14\% of active users), or posted images of damages in their locality (20.1\%), posted their opinion on issues related to Amphan (13.4\%), and so on.
An important usage of OSM seems to be that a sizable fraction of people are using OSM to inquire about safety of others (27.9\% of the active users) and to inform others about their own safety (26.8\%). The use of OSM for inquiring about safety of others and informing others of one's own safety has been observed during other disaster events as well~\cite{osm-emergencies-15yrs}.
%\footnote{\url{https://www.telegraph.co.uk/technology/information-age/how-social-media-can-help-during-crisis/}}
Around 15.6\% people also posted information to help others.
A small fraction of participants (5.6\%) specifically connected to social media accounts of the Government agencies or service providers and asked for help.

\vspace{2mm}
\noindent \textbf{Useful information received from social media:}
We then asked {\it what kind of useful information they received from OSM related to Amphan}. The responses are summarized in Table~\ref{tab6:socialinfo}. 
Among the participants who used social media within 7 days after Amphan, 18.4\% did not get any useful information related to Amphan.
The rest 81\% of participants received various useful information, including 33\% of them receiving useful updates from government and/or service providers. 
As many as 43\% of active participants received information about safety of others, 31.3\% received important updates from their own locality, and 49.2\% received important updates from other regions.
We also asked the participants to briefly describe some social media posts which they felt contained important information on Amphan. 
Some of the responses by the participants are shown in Table~\ref{tab:exampleposts}. From this table, we observe another important usage of OSM in organizing online donation campaigns. 

\vspace{3mm}
\noindent These numbers show that OSM are more popularly used to consume information than produce information. 
Also, it can be noted that the engagement by the authorities is quite low after the cyclone, as compared to before the cyclone --  while 62.7\% of all the survey participants received advance warning {\it before} the cyclone via OSM, only 29.4\% received useful updates from authorities in the 1-week period {\it after} the cyclone.
This big difference between the two fractions shows that Government and other authorities can use social media to reach out to more people after a disaster, as shown also shown in studies~\cite{10.1145/3137597.3137602, 10.1145/2771588}. 
Especially, online social media is even more beneficial since the authorities can do both -- broadcast information to everyone, and engage with personal user-accounts individually -- which is not possible with traditional media like television or newspapers.

\vspace{2mm}
\noindent \textbf{Harmful/undesirable information received from social media:}
Finally, we also asked if the participants observed any undesirable / harmful content on social media, which are known to often circulate on social media after a crisis~\cite{10.1145/3161603,disaster-fake-news-osm}. 
%(as explored by Zubiaga et.al.~\cite{10.1145/3161603}), and spread panic among people. 
The responses to our question are summarized in Table~\ref{tab6:socialnegative}.
Among the participants who used social media within 7 days after Amphan, 48\% said they observed some undesirable / harmful posts -- 
19.6\% said they observed fake news/rumors, 21.2\% said they saw too much useless posts, 34.1\% observed too much of political arguments, and 7.8\% saw too much of superstitious/religious posts. 
We also asked the participants to briefly describe some social media posts that they thought had a negative impact on the situation. Some of their responses are shown in Table~\ref{tab:examplenegative}.

\subsection*{Summary of Section}

\noindent As discussed earlier, our sample of survey participants is biased towards frequent social media users; but this bias actually makes the analysis in this section more credible. 
%We found that most people are consumers of information from social media than producers. 
We found that people use OSM in many different ways in the aftermath of the disaster.
Importantly, out of the `active' survey participants who used OSM during the 7 days immediately after Amphan,
81.6\% said they received some useful information related to Amphan from OSM, while 52\% said they observed undesirable/harmful content on OSM. Hence, {\it a much larger fraction of users (who actively use social media) seem to find useful information via OSM, than the fraction of users who find OSM propagating harmful content}.

Also, our study indicates that Government agencies and service providers should engage more with people via OSM in the aftermath of a disaster, to give them localised and personalized updates on the situation.

\section{Preparedness of Authorities}
\label{authorities}

In this section we study how vigilant the authorities (including government and private service providers) were to mitigate the damages caused by the cyclone. 
%We analyse the ratings that the participants had given to the preparedness of the authorities, and try to find which were the important factors for their score.

\vspace{2mm}
\noindent \textbf{Warning people about the cyclone in advance:}
We asked the participants if they had received advance warning about the cyclone, and if so, via which medium. The responses are given in Table~\ref{tab6:warning}. 
The authorities were effective in informing a large majority of people via news media (94.5\%), but relatively less through direct communication (31.3\%). A large fraction (62.7\%)  also received warning from social media.

In fact, one of our participants from a remote region of West Bengal commented -- ``The Govt. of West Bengal warned us through SMS that those who were not staying in {\it pukka} houses should find some place safe and stay there for the next few days. It also specifically warned residents near coastal areas.''
Hence, the authorities were very much effective in warning people about the cyclone.

\begin{table}[hbt]
    \centering
    \begin{tabular}{|l|c|}
        \hline
        \bf Medium of communication & \bf \% \\
        \hline \hline
        Did not receive any warning & 2.0\% \\
        Via communication from Government (SMS/phone/email) & 31.3\% \\
        Via announcements on news media (TV, newspaper) & 94.5\% \\
        Via word of mouth from friends and relatives & 36.8\% \\
        Via social media (Facebook/Twitter/Whatsapp) & 62.7\% \\
        \hline
    \end{tabular}
    \caption{\bf Communication medium through which our participants received warning  about the cyclone.}
    \label{tab6:warning}
\end{table}

\vspace{2mm}
\noindent \textbf{Responses received from authorities after the cyclone:} 
We asked the participants if they had tried to  contact any of the authorities for assistance immediately after the cyclone. About half of the participants said that they did {\it not} attempt to contact any authority. 
To the other half of the participants (who tried to contact authorities for help), we asked whether they received helpful responses from the authorities. This is the distribution of their responses:
\begin{itemize}
    \item Could not contact authorities (or no response): 53\%
    \item Could contact but did not receive helpful response: 12\%
    \item Could contact and received helpful response: 35\%
\end{itemize}
It is alarming that as much as 65\% of the participants who tried to contact authorities could not actually contact them or did not get any helpful response. 
This high fraction may be because (i)~the cyclone damaged the helplines as well, and (ii)~due to the ongoing lockdown situation, the authorities could deploy a reduced number of personnel (compared to non-pandemic situations).
%However, it is reassuring that most people who could contact the authorities received helpful responses.
Overall, these statistics show that authorities should improve their responsiveness after a disaster.

\begin{figure}[hbt]
    \centering
    \includegraphics[width =0.9\linewidth,height=5cm]{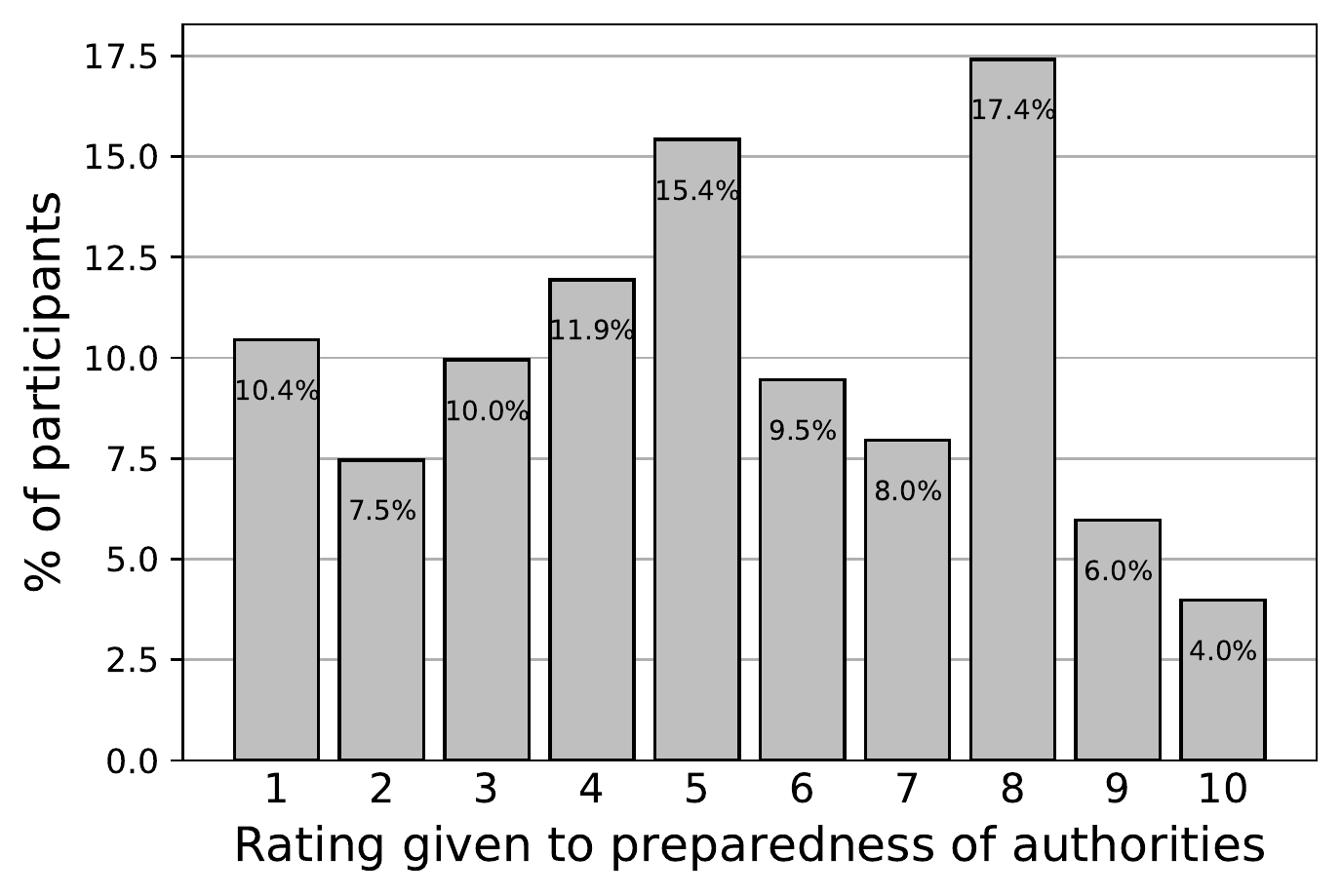}
    \caption{\bf Preparedness of authorities as perceived by our participants.}
    \label{fig7:preparedness}
\end{figure}

\begin{table*}[hbt]
    \centering
    \normalsize
    \begin{tabular}{|l|c|c|c|c|c|c|}
        \hline
        \shortstack[l]{\bf Participants who said authorities were\\ \:} & \shortstack{\bf N24P\\(60)} & \shortstack{\bf KOL\\(59)} & \shortstack{\bf S24P\\(22)} & \shortstack{\bf HGLY\\(18)} & \shortstack{\bf HWH\\(14)} & \shortstack{\bf Others\\(28)} \\
        \hline \hline 
         Highly Prepared (Rating $\ge 8$)& 40.0\% & 20.3\% & 18.2\% & 11.1\% & 21.4\% & 35.7\% \\
         Poorly Prepared (Rating $\le 3$) & 23.3\% & 37.3\% & 27.3\% & 27.8\% & 28.6\% & 17.9\% \\
%        There was Agitation in the locality & 28.3\% & 32.2\% & 63.6\% & 44.4\% & 57.1\% & 10.7\% \\
        \hline
    \end{tabular}
    \caption{\bf Rating given to authorities (eg, government, service providers) on their preparedness. The numbers in brackets below each of the headers represent the actual count of participants from that district.}
    \label{tab7:districtpreparedness}
\end{table*}

\vspace{2mm}
\noindent \textbf{Rating the preparedness of authorities:} Finally, we asked the participants to rate the preparedness of the authorities serving their localities on a Likert scale of 1-10, where level `1' means ``authorities were not prepared at all'' and level `10' means ``authorities were excellently prepared''.

The distribution of the ratings by our survey participants is shown in Figure~\ref{fig7:preparedness}. 
The mean preparedness rating (averaged across all participants) is $5.25$, while the median preparedness rating (across all participants) is $5.0$.

We consider the authorities to be \textit{highly prepared} (from the point of view of a participant) if they were rated greater than or equal to 8.
On the other hand, we consider the authorities to be \textit{poorly prepared} (from the point of view of a participant) if they were rated less than or equal to 3. 
Out of the 201 participants, almost an equal number  said authorities were highly prepared (55 participants) and authorities were poorly prepared (56 participants).

%We considered the two sets of participants mentioned above -- those who said authorities were highly prepared, and those who said authorities were poorly prepared -- and computed the district-wise distribution of the two sets of participants. 
Table~\ref{tab7:districtpreparedness} shows the district-wise fractions of participants (i)~who said authorities were highly prepared, and (ii)~who said authorities were poorly prepared.  
Interestingly, across all districts, there are both participants who said authorities were highly prepared, as well as participants who said authorities were poorly prepared.
The authorities seem to have been best prepared in North 24 Paraganas -- 40\% of the participants from this district said authorities were highly prepared, and much fewer participants (23.3\%) said authorities were poorly prepared. 
On the contrary, in Kolkata, only 20.3\% said authorities were highly prepared but 37.3\% said authorities were poorly prepared.
Apart from North 24 Paraganas and `Others' (districts that were not affected much by the cyclone), in all other districts, a larger fraction of participants felt the authorities were poorly prepared (as compared to the fraction who felt authorities were highly prepared).

Note that the preparedness of the authorities was most probably hampered by the ongoing COVID19-induced lockdown which made it difficult for the authorities to deploy sufficient manpower in many regions.

%We had also asked the participants if there was agitation in their locality regarding the disruption of services. Percentage wise it seems most of the agitation was from South 24 Paraganas and Howrah, however the support is less (14 and 18 respectively). Count wise Kolkata was highest (28). This correlates with the unavailability of services for long duration of time.

\begin{table}[tb]
    \centering
    \begin{tabular}{|l|c|c|}
        \hline
        Among those who tried to & \multicolumn{2}{|c|}{People saying authorities were}\\
        \cline{2-3}
        contact authorities &  Highly Prepared & Poorly Prepared\\
        \hline
        Total Count & 25 & 36\\
        \hline
        Received helpful response & \blue{72.0\%} & 11.1\%\\
        Received no/unhelpful response & 28\% & \red{88.9\%} \\
        \hline
    \end{tabular}
    \caption{\bf Response received from authorities vs rating given to preparedness of authorities. Total count consists of only the participants who tried to contact the authorities.}
    \label{tab7:responsecorr}
\end{table}

\begin{table}[tb]
    \centering
    \begin{tabular}{|l|c|c|}
        \hline
        Among those facing disruption, & \multicolumn{2}{|c|}{People saying authorities were}\\
        \cline{2-3}
        electricity supply was restored in &  Highly Prepared & Poorly Prepared\\
        \hline
        Total Count & 49 & 51\\
        \hline
        Less than 1 day & \blue{57.2\%} & 45.1\% \\        
        Between 1-2 days & 16.3\% & 5.9\% \\
        More than 2 days & 26.5\% & \red{49.0\%} \\
        \hline
    \end{tabular}
    \caption{\bf Time for restoration of electricity supply vs rating given to preparedness of authorities. Total count consists of only the participants who faced disruption in electricity supply.}
    \label{tab7:electricitycorr}
\end{table}

\vspace{2mm}
\noindent \textbf{Reasons for good and bad ratings:} 
We wanted to identify some probable factors that could have affected the preparedness ratings given by participants. Obviously, the actual reason behind a participant's rating can only be found through personal interaction. Since such interaction was not feasible, we checked if some specific factors are correlated with the ratings given by participants.

Our hypothesis is that a person's rating of the preparedness of authorities can be affected by factors such as whether he/she received helpful responses from the authorities, how long he/she had to tolerate disruptions in electricity/phone/internet services, and so on. We now investigate these hypotheses. 

%We have tried to find out the reasons for participants rating the authorities to be highly or poorly prepared. 
Table~\ref{tab7:responsecorr} shows the correlation between the preparedness rating given by a participant, and whether the said participant received helpful responses from the authorities. 
We see that 72\% of the participants who received helpful responses said that the authorities were highly prepared, while 88.9\% of those who did not receive any helpful response said that the authorities were poorly prepared.
These figures seem to corroborate our hypothesis that participants who received helpful responses would tend to say authorities were highly prepared, whereas those who did not receive helpful responses, would tend to say that the authorities were poorly prepared. 

Similarly, Table~\ref{tab7:electricitycorr}, Table~\ref{tab7:phonecorr} and Table~\ref{tab7:internetcorr} respectively show
the correlation between the preparedness ratings given by participants, and the time taken for restoration of electricity supply, phone services, and internet services for those participants. 
All the three tables shown a common trend (that agrees to our hypothesis) -- people who faced disruption of these services for less than 1 day mostly opined that the authorities were highly prepared (fractions shown in blue-colored text). Whereas, people facing disruption of these services for more than 2 days have mostly said authorities were poorly prepared (fractions shown in red-colored text).

\begin{table}[tb]
    \centering
    \begin{tabular}{|l|c|c|}
        \hline
        Among those facing disruption, & \multicolumn{2}{|c|}{People saying authorities were}\\
        \cline{2-3}
        phone connection was restored in &  Highly Prepared & Poorly Prepared\\
        \hline
        Total Count & 44 & 49\\
        \hline
        Less than 1 day & \blue{45.5\%} & 20.4\% \\
        Between 1-2 days & 29.5\% & 8.2\% \\
        More than 2 days & 25.0\% & \red{71.4\%} \\
        \hline
    \end{tabular}
    \caption{\bf Time for restoration of phone connection vs rating given to preparedness of authorities. Total count consists of only the participants who faced disruption in phone services.}
    \label{tab7:phonecorr}
\end{table}

\begin{table}[tb]
    \centering
    \begin{tabular}{|l|c|c|}
        \hline
        Among those facing disruption, & \multicolumn{2}{|c|}{People saying authorities were}\\
        \cline{2-3}
        internet connection was restored in &  Highly Prepared & Poorly Prepared\\
        \hline
        Total Count & 46 & 50\\
        \hline
        Less than 1 day & \blue{30.4\%} & 8.0\% \\
        Between 1-2 days & 17.4\% & 4.0\% \\
        More than 2 days & 52.2\% & \red{88.0\%} \\
        \hline
    \end{tabular}
    \caption{\bf Time for restoration of internet connection vs rating given to preparedness of authorities. Total count consists of only the participants who faced disruption in internet services.}
    \label{tab7:internetcorr}
\end{table}

\subsection*{Summary of Section}

\noindent While the authorities did a very creditable job of warning people of the cyclone in advance, the performance of the authorities {\it after} the cyclone was not equally impressive.
A large majority of people who tried to contact authorities for help after the cyclone either could {\it not} contact the authorities, or did not get helpful responses from the authorities.
Regarding preparedness of authorities, an equal number of participants rated the authorities to be highly prepared and poorly prepared to deal with the effects of the cyclone. 
Notably, in four out of the five districts that were majorly affected by cyclone Amphan, more people judged the authorities to be poorly prepared.
These overall poor preparedness ratings of authorities seem to be correlated with factors such as whether participants received helpful responses, or how long it took for restoration of essential services. 
%Most people who have have received helpful response have rated the authorities as highly prepared, and those who did not rated the opposite. People also rated based on the delay in restoration of services
\section{Implications}\label{sec:implications}

\noindent Finally, in this section we present our synthesis on how, taking lessons from this disaster, the authorities can better handle future disasters. 
Specifically, we present our insights across two fronts. 
First, as part of the survey, we asked the participants -- ``{\it If you have  any  suggestions  that  will  help  to  better  prepare  for  future  disaster  events, please list them here}''; we will present our findings from the participant's responses to this question. 
Second, we will discuss how our survey results strongly indicate the efficacy (as well as pitfalls) of using social media to help people in the times of disaster.

\begin{table*}[hbt]
\centering
\begin{tabular}{|p{0.1\textwidth}|p{0.85\textwidth}|}
    \hline
    \textbf{Themes} & \textbf{Responses from participants} \\
    \hline \hline
    \multirow{4}{=}{Uprooting of trees} & - Army must be kept ready to take care of the cleaning of uprooted trees. \\
            & - The trees in localities could be trimmed before so that uprooting can be avoided.\\
            & - Trees are not regularly cut and maintained. No pruning is done even before the monsoons. If regularly cut and pruned they would, even if uprooted, caused less damage.\\
    \hline \hline
    \multirow{10}{=}{Electricity, mobile, Internet supplies} & - Electricity dept should be more adequate and engage more people when the workload is so huge. \\
            & - Have power backup generators ready. \\
            & - Using underground cables for power distribution in clustered places like here in Kolkata. Using underground optical fibre for data transmission. \\
            & - Targeting electricity to mobile towers and communication mediums so that news can travel and authorities can be contacted and people can be helped out. \\
            & - Before restoring electricity supply, it should be checked whether the locality is still waterlogged. The waterlogging resulted in "electric fires" on several electricity poles when power supply was restored. \\
            & - If main line electric supply wire take down from electric post before storm came then electric supply does not damage that much\\
    \hline \hline
    \multirow{4}{=}{Drinking water and food} & - I think proper drinking water should be kept in store from beforehand in adequate amount, so that it can be supplied to people lacking drinking water, later in a systematic manner.\\
            & - Those who are severely affected by the cyclone need to be served with food for at least a few days.\\
            & - Drinking water tankers should be ready to supply after disaster. \\
    \hline \hline
    \multirow{12}{=}{General disaster response} & - Frequent mock drills within the disaster response teams and frequent maintenance of the tools/machines used by them. \\
            & - Need to have synchronised effort all the related departments and Govt Agencies\\
            & - The Corporation, electric supply and other authorities should be well prepared before with lots of manpower and equipment for disaster. \\
            & - It'll be better that if we can provide an idea to those people who are staying in village about the severeness of the storm on the basis of the estimated speed, so that they can take preparation in a much better way. Not only do's. Tell the reason why to do.. \\
            & - Improve drainage system, increase forest more specifically Sundarban, make concrete structures in safe distance from the sea. \\
            & - When is such natural calamity is forecasted, I feel the roadside dwellers or homeless people and animals must be shifted to nearby schools, colleges, or any vacant and safe buildings so that they don't face the nature's wrath and be safe. \\
            & - A toll free number should be given prior to such natural calamity and a quick response team must be on standby in every locality \\
    \hline
\end{tabular}
\caption{\bf Sample responses (excerpts) by participants when asked to give ``suggestions that will help to better prepare for future disaster events''.}
\label{tab:examplesuggestions}
\end{table*}

\subsection{Suggestions to better prepare for future disasters}

\noindent We identify a few key themes from the general responses of the participants when asked to give suggestions  that  will  help  to  better  prepare  for  future  disaster  events. We present those themes as well as a few verbatim user responses in Table~\ref{tab:examplesuggestions}. Specifically, the participants expressed ideas to better handle future disasters in terms of the after-effects. 

\vspace{2mm}
\noindent \textbf{Continuous maintenance}: Our participants pointed out ways to minimize damage due to uprooting of trees via continuous maintenance (e.g., cutting and trimming the trees regularly). 

\vspace{2mm}
\noindent \textbf{Advance planning to employ more resources}: The participants opined that, since a major disaster like Amphan can be expected to disrupt services such as electricity, mobile and internet, the authorities need to have a better `preparedness planning' so that more manpower and backup infrastructures can be quickly called upon to ensure functioning of these services. For instance, power back generators can be kept ready, and army personnel can be kept prepared to help with cleaning uprooted trees.
In fact, similar suggestions also came up while solving the problem with lack of drinking water and food in certain areas (e.g., ``\textit{water  should  be  kept  in  store  from  beforehand  in  adequate  amount}'') 

\vspace{2mm}
\noindent \textbf{Improving co-ordination}: An important point that was mentioned was the improvement of co-ordination among various authorities (e.g., between municipality and electricity supplier), e.g., ``{\it Need to have synchronised effort all the related departments and Govt Agencies}''. It seems that lack of such co-ordination really hindered the efforts of the authorities. 
For example one participant mentioned that ``\textit{Before restoring electricity supply, it should be checked whether the locality is still waterlogged. The waterlogging resulted in ``electric fires'' on several electricity poles when power supply was restored.}'' So, more information sharing between different authorities (e.g., those who are helping with solving water-logging and those who are fixing affected electrical lines) in crucial for a swifter response. 

\vspace{3mm}
\noindent We note that most of the suggestions discussed above were a direct response (from the participants) to the situation at hand, and aimed at addressing the specific problems faced by them in-spite of the best efforts of authorities. 
Additionally, we further analyzed communication between the authorities and citizens via novel mediums like social media--an avenue which helped some people to avert the worst in this disaster and could be further improved by developing more systematic mechanisms for communication. 

\subsection{Utility of social media in post-disaster scenario}

\noindent We note in Table~\ref{tab:examplesuggestions} that some participants pointed out the value of communication between authorities and affected populace. 
One participant mentioned that \textit{``A toll free number should be given prior to such natural calamity and a quick response team must be on standby in every locality''}. 
However, given the huge spread of the affected area and the rarity of such disaster events, such a centralized approach might be too costly to build for authorities. 
To that end, our data however, point out a few very interesting observations on how to improve the efficacy of two-way communication between authorities and affected citizens.

\vspace{2mm}
\noindent \textbf{The increasing reach of social media}: Recall that Table~\ref{tab6:warning} identified that traditional broadcast mediums such as newspaper and television fared the best in warning people of the impending disaster -- 94\% of our participants received advance warning through these mass communication mediums. 
However, these mediums have two problems -- they are costly, and they provide one-way communication (i.e., they do not provide any simple way for the common people to reach the authorities with their questions or concerns). 
On the other hand, online social media (OSM) can reach the populace at a minimal fraction of the cost, and also provides a simple way for citizens to reach authorities (i.e., two-way communication). 
One key finding from Table~\ref{tab6:warning} is that, as many as 62.7\% of the participants received warning via OSM, which is second only to the mass communication mediums. 
Additionally, even though Amphan led to severe disruption of Internet connectivity, more than 80\% of our survey participants received useful information about Amphan via social media in the $7$ days after the cyclone (as seen from Table~\ref{tab6:socialinfo}).
In fact, a major utility of social media seems to be in inquiring about safety of friends/relatives and informing others about one's own safety (as reported by 30\%--40\% of the participants).
These numbers underline the increasing outreach of this medium, at least among the urban population who are conversant with use of such novel mediums.

\vspace{2mm}
\noindent \textbf{Lowering the cost of disseminating and receiving information}: OSM provide a {\it low-cost} and {\it real-time} medium for the authorities to reach the citizens, as well as for the citizens to express their worries to the right people. 
%So OSM can really help the authorities to reach a vast number of people in real-time. 
Furthermore, OSM can often help (due to the low cost and ubiquity of mobile Internet) to receive information about the affected remote places which are often {\it not covered} by traditional mediums. 
For instance, one participant mentioned that \textit{``I feel that the (social media) posts, stating the devastating condition of worst-hit area Sundarban (a remote region in West Bengal) due to Amphan are helpful in reality. Many can contribute to the funds which are shared through social media''}. So, our results identify that developing mechanisms to timely disseminate information on easily-accessible social media can tremendously help the affected populace to cope up with their loss.

It can be noted that, recent research works have developed mechanisms for using digital media in spite of disruptions in Internet connectivity~\cite{disaster-internet-blackout,surakshit}. 
Such mechanisms can be set up during major disasters to ensure connectivity among the people, which will increase the access to social media and other digital media in post-disaster scenarios.

\vspace{2mm}
\noindent \textbf{Enabling citizens to get responses from authorities}: We note from Table~\ref{tab7:responsecorr} that the event of receiving responses to Amphan-related queries is highly positively correlated with the favorable perception of people about preparedness of authorities. 
However, it is tremendously hard for people to receive responses to their queries via traditional mechanisms like phone. 
To that end, we propose that our survey results demonstrate a need for authorities to be more proactive in reaching out and answering queries to help the affected populace via digital media (including social media). 
To this end, novel Artificial Intelligence (AI)-driven chatbots may provide a scalable and automated way to pave this two-way communication during or after future disasters.

\vspace{2mm}
\noindent \textbf{Facilitating trustworthy and useful news on OSMs}: 
OSM have a severe pitfall today --- often, they are {\it not} a reliable medium to connect with general populace due to misinformation, abuse and in general, lack of trustworthy information~\cite{disaster-fake-news-osm}. 
Additionally, there is often lot of `conversational chatter' on OSM (such as political propaganda) which does not help post-disaster relief operations, and in fact, obscures the critical information.
Table~\ref{tab:examplenegative} identified some types of misinformation that worsen the situation. 
Often such misinformation (e.g., rumors) is just a side-effect of lack of good, authentic information.
Thus, we strongly prescribe the authorities to disseminate more useful information from trustworthy official accounts to drown such misinformation. 
Additionally, recent AI-based techniques can be employed to extract critical information from social media that would help post-disaster relief operations~\cite{tcss_Basu,ipm_Dutt,rudra-cikm-disaster}, as well as to identify and counter rumors and other misinformation~\cite{disaster-fake-news-osm,RudraSGG18}.
Thus, creating AI-based automated systems to aggregate useful information from OSM and disseminate trustworthy information on OSM will quite possibly help people to better cope with natural disasters in future.

\section{Conclusion} \label{sec:conclu}

In this report, we analysed the responses to an online survey (by 201 participants) to get insights about the damages caused by the super-cyclone Amphan, and its impact on people. 
We saw that our participants were severely affected by disruption of internet, phone and electric services. In fact, more than 40\% of our participants responded that these services took more than 4 days to be restored. 
We also found that people have rated the authorities highly if they were responsive and vigilant enough to repair the damages caused within a day. 
Furthermore, our exploration revealed that communication (or lack there-of) between the authorities and affected citizens strongly influenced public perceptions about preparedness of the authorities. 
Then we synthesized our findings into policy implications for better handling future disasters. 
The survey results strongly suggest that authorities can better handle such situations using social media and AI-driven techniques over digital mediums that are low-cost, personalized alternatives of traditional communication mediums. 

As stated earlier, some of the observations from the survey can be affected by the ongoing COVID19-induced lockdown situation, as well as by the bias of our survey participants towards students residing in urban areas.
As a result, some of the micro-level findings might not generalize in another sample and another time. 
However, we believe that the broad findings (e.g., communication from authorities is a necessary aspect, how people use social media during a disaster, what affects people the most) should generalize across various disaster events.

We believe that this work will pave the way forward in better understanding of the impact of disasters like Amphan on the general populace, and also help to build novel mechanisms for authorities to cope with the after-effects of such disasters.

\section*{Acknowledgements}
%\vspace{5mm}
%\noindent {\bf Acknowledgemnts:} 
We sincerely thank all the participants who gave their valuable time in responding to the survey. 
We also acknowledge Abhisek Dash, Shalmoli Ghosh, Shounak Paul and Paheli Bhattacharya (all from IIT Kharagpur) for their constructive suggestions and help in designing the survey.
We especially thank Moumita Basu (of UEM Kolkata) for her insights on the survey design, as well as for her valuable help in disseminating the survey widely.
This research is partially funded by the Sponsored Research and Industrial Consultancy (SRIC) unit, IIT Kharagpur, through the project ``Building Information Systems for Emergency Relief and Preparedness''.

\nocite{*}
\bibliographystyle{ieeetr}
\bibliography{ref}

\appendix

%\subsection{Survey Questions} \label{app:survey1instrument}

\noindent We state below the exact questions asked in the survey.

\vspace{5mm}

\begin{scriptsize}

\noindent \textbf{Some details about you and your locality}

\noindent In this survey, you will be asked questions about the effect of cyclone Amphan on your LOCALITY. For purpose of this survey, consider your "locality" to be the region within approx. 500 meters from your home in all directions.  \\[4pt]

\noindent PIN code of your locality: \_\_\_.\\

\noindent A more specific identifier for your locality (e.g., street name, specific name of area): \_\_\_ \\

\noindent Were you yourself in the said locality during the cyclone, or are you filling in the survey on behalf of someone else who was in that locality (e.g., your family or relatives)? I was myself in the said locality during cyclone Amphan $\bigcirc$ I am filling in the survey on behalf of someone else who was in that locality $\bigcirc$ Other \_\_\_\\

\noindent What is your age group? $<20$ years $\bigcirc$ 21-30 years $\bigcirc$ 31-40 years $\bigcirc$ 41-50 years $\bigcirc$ $>50$ years $\bigcirc$ Would prefer not to say $\bigcirc$  \\

\noindent What is your gender? Male $\bigcirc$ Female $\bigcirc$ Other $\bigcirc$ Would prefer not to say $\bigcirc$ \\

\noindent What is your occupation? (E.g., student, teacher, businessman, doctor, engineer, software professional, ...) \_\_\_\\

\noindent \textbf{Before the cyclone}

\noindent How did you receive advance warning about cyclone Amphan? You can select one or more options. 
Did not receive any advance warning $\bigcirc$ 
Via announcements on news media (TV / newspaper / radio) $\bigcirc$ 
Via social media (e.g., Facebook / Twitter / WhatsApp) $\bigcirc$ 
Via word of mouth from friends / relatives $\bigcirc$ 
Via specific communication from Government (SMS / WhatsApp / phone / email, etc.) $\bigcirc$ 
Others $\bigcirc$ \\

\noindent Were any services disrupted even BEFORE cyclone Amphan struck your locality? You can select one or more options. No service was disrupted before the cyclone  $\bigcirc$ Electricity was disrupted  $\bigcirc$ Mobile connectivity was disrupted  $\bigcirc$ \\

\noindent Were you asked to evacuate your locality before the cyclone struck? Yes  $\bigcirc$ No  $\bigcirc$ Cannot say  $\bigcirc$ \\

\noindent \textbf{Damage in your locality due to Amphan}

\noindent Approximately how many trees were uprooted / seriously damaged in your locality due to Amphan? Cannot say $\bigcirc$ No tree was uprooted / seriously damaged in my locality $\bigcirc$ Between 1 and 5 trees $\bigcirc$ Between 6 and 10 trees $\bigcirc$ More than 10 trees  $\bigcirc$\\

\noindent Approximately how many buildings were damaged in your locality due to Amphan? Assume "damaged" means anything more severe than breaking of glass windows. Cannot say $\bigcirc$ No building was damaged in my locality $\bigcirc$ Between 1 and 5 buildings $\bigcirc$ Between 6 and 10 buildings $\bigcirc$ More than 10 buildings  $\bigcirc$\\

\noindent Approximately for how long was your locality waterlogged / flooded due to Amphan? My locality was not waterlogged $\bigcirc$ Less than 6 hours $\bigcirc$ Between 6 hours and 12 hours $\bigcirc$ Between 12 hours and 1 day $\bigcirc$ Between 1 day and 2 days $\bigcirc$  Between 2 days and 3 days $\bigcirc$ Between 3 days and 4 days $\bigcirc$ More than 4 days $\bigcirc$ My locality is still waterlogged $\bigcirc$\\

\noindent Approximately for how long was drinking water supply irregular in your locality? Drinking water supply was not affected in my locality $\bigcirc$ Less than 6 hours $\bigcirc$ Between 6 hours and 12 hours $\bigcirc$ Between 12 hours and 1 day $\bigcirc$ Between 1 day and 2 days $\bigcirc$  Between 2 days and 3 days $\bigcirc$ Between 3 days and 4 days $\bigcirc$ More than 4 days $\bigcirc$ Drinking water supply not regular till now $\bigcirc$\\

\noindent \textbf{Disruption of Electricity}

\noindent Which electricity supplier are you served by? CESC $\bigcirc$ WBSEDCL / WBSEB $\bigcirc$ Other $\bigcirc$\\

\noindent Was electricity disrupted in your locality due to cyclone Amphan? Yes $\bigcirc$ No $\bigcirc$ Cannot say $\bigcirc$ \\

\noindent Approximately how long after the cyclone did normal electricity service resume in your locality? Within 6 hours $\bigcirc$ Between 6 hours and 12 hours $\bigcirc$ Between 12 hours and 1 day $\bigcirc$ Between 1 day and 2 days $\bigcirc$  Between 2 days and 3 days $\bigcirc$ Between 3 days and 4 days $\bigcirc$ More than 4 days $\bigcirc$ Normal electricity service not resumed till now $\bigcirc$ \\

\noindent Were electric poles / cables / transformers damaged in your locality due to the cyclone? Yes $\bigcirc$ No $\bigcirc$ Cannot say $\bigcirc$ \\

\noindent Are there any PUBLIC UTILITIES in your locality, whose operations were hampered due to disruption in electricity? You can choose one or more options. No public utility in my locality $\bigcirc$ Hospital / nursing home $\bigcirc$ Water pumping station $\bigcirc$ Mobile tower $\bigcirc$ Other $\bigcirc$ \\

\noindent \textbf{Disruption in phone connectivity}

\noindent Which phone service provider do you use? You can select more than one option, e.g., if you or your family members use multiple service providers. Airtel mobile $\bigcirc$ BSNL mobile $\bigcirc$ Jio mobile $\bigcirc$ Vodafone mobile $\bigcirc$ Landline phone $\bigcirc$ Other $\bigcirc$\\

\noindent Was phone connectivity disrupted in your locality due to Amphan? Yes $\bigcirc$ No $\bigcirc$ Cannot say $\bigcirc$ \\

\noindent Which of these phone services were disrupted? You can select one or more options. Airtel mobile $\bigcirc$ BSNL mobile $\bigcirc$ Jio mobile $\bigcirc$ Vodafone mobile $\bigcirc$ Landline phone $\bigcirc$ Other $\bigcirc$\\

\noindent Approximately how long after the cyclone did normal phone connectivity resume in your locality?  [In case different services resumed at different times, please specify for that phone service which resumed the earliest] Within 6 hours $\bigcirc$ Between 6 hours and 12 hours $\bigcirc$ Between 12 hours and 1 day $\bigcirc$ Between 1 day and 2 days $\bigcirc$  Between 2 days and 3 days $\bigcirc$ Between 3 days and 4 days $\bigcirc$ More than 4 days $\bigcirc$ Normal phone connectivity not resumed till now $\bigcirc$ \\

\noindent \textbf{Disruption in internet connectivity}

\noindent How do you connect to the internet? You can select one or more options. Through mobile phone $\bigcirc$ Through landline  $\bigcirc$ Cable internet $\bigcirc$ Other $\bigcirc$\\

\noindent Was internet connectivity disrupted in your locality due to Amphan? Yes $\bigcirc$ No $\bigcirc$ Cannot say $\bigcirc$ \\

\noindent Approximately how long after the cyclone did normal internet connectivity resume in your location? Within 6 hours $\bigcirc$ Between 6 hours and 12 hours $\bigcirc$ Between 12 hours and 1 day $\bigcirc$ Between 1 day and 2 days $\bigcirc$  Between 2 days and 3 days $\bigcirc$ Between 3 days and 4 days $\bigcirc$ More than 4 days $\bigcirc$ Normal internet connectivity not resumed till now $\bigcirc$\\

\noindent \textbf{How much were YOU / YOUR FAMILY affected by the cyclone?}

\noindent We are stating below some possible aspects of damage. In the context of each aspect, please rate how much you or your family were affected in the scale of 1-5. Here 1 indicates "I / my family was not affected at all" and 5 indicates "I / my family was extremely severely affected ".\\

\noindent Uprooting of trees: 
1 $\bigcirc$ 2 $\bigcirc$ 3 $\bigcirc$ 4 $\bigcirc$ 5 $\bigcirc$ \\

\noindent Damage to buildings: 
1 $\bigcirc$ 2 $\bigcirc$ 3 $\bigcirc$ 4 $\bigcirc$ 5 $\bigcirc$ \\

\noindent Waterlogging / flooding: 
1 $\bigcirc$ 2 $\bigcirc$ 3 $\bigcirc$ 4 $\bigcirc$ 5 $\bigcirc$ \\

\noindent Disruption of electric supply: 
1 $\bigcirc$ 2 $\bigcirc$ 3 $\bigcirc$ 4 $\bigcirc$ 5 $\bigcirc$ \\

\noindent Disruption of phone connectivity: 
1 $\bigcirc$ 2 $\bigcirc$ 3 $\bigcirc$ 4 $\bigcirc$ 5 $\bigcirc$ \\

\noindent Disruption of internet connectivity: 
1 $\bigcirc$ 2 $\bigcirc$ 3 $\bigcirc$ 4 $\bigcirc$ 5 $\bigcirc$ \\

\noindent Unavailability of drinking water:
1 $\bigcirc$ 2 $\bigcirc$ 3 $\bigcirc$ 4 $\bigcirc$ 5 $\bigcirc$ \\

\noindent \textbf{Other questions about your locality}

\noindent Was there any agitation in or around your locality after the Amphan cyclone, to protest against non-availability of essential supplies (e.g., electricity, water)? Yes $\bigcirc$ No $\bigcirc$ Cannot say $\bigcirc$ \\

\noindent Which of the following is true regarding waterlogging / flooding in your locality?\\
$\bigcirc$ My locality gets regularly flooded during heavy rains, and was also flooded due to Amphan\\ 
$\bigcirc$ My locality does NOT usually get flooded, but was flooded due to Amphan\\
$\bigcirc$ My locality did not get flooded due to Amphan

\noindent If you or your locality faced any other problems due to the cyclone Amphan, please list them here. \_\_\_\\

\noindent \textbf{Use of social media}

\noindent NORMALLY, how frequently do you use social media (i.e., visit sites like Facebook, Twitter, WhatsApp, Instagram, etc.)? Almost always online $\bigcirc$ A few times a day $\bigcirc$ A few times a week $\bigcirc$ A few times a month $\bigcirc$ I do not use social media $\bigcirc$ \\

\noindent In general, which social media sites do you use? You can select one or more options. Twitter $\bigcirc$ WhatsApp $\bigcirc$ Facebook $\bigcirc$ Instagram $\bigcirc$ Other \_\_\_\\

\noindent Have you POSTED any information related to Amphan on social media, within 7 days after the cyclone? You can select one or more options. \\
$\bigcirc$ I did not use social media within 7 days after Amphan\\
$\bigcirc$ I used social media, but did not post any information related to Amphan\\
$\bigcirc$ Generally described the situation in my locality\\
$\bigcirc$ Posted images of the damages in my locality\\
$\bigcirc$ Connected to Government accounts / NGOs / service providers and asked for help\\
$\bigcirc$ Inquired about safety of others\\
$\bigcirc$ Posted information that can help others (e.g., helpline numbers, assured safety of others in my region)\\
$\bigcirc$ Informed others about safety of myself / my family\\
$\bigcirc$ Posted my opinion on issues related to Amphan\\
$\bigcirc$ Other \_\_\_\\

\noindent Have you RECEIVED any useful information related to Amphan from social media, within 7 days after the cyclone? You can select one or more options.

\noindent$\bigcirc$ I did not use social media within 7 days after Amphan\\
$\bigcirc$ I used social media, but did not receive any useful information related to Amphan\\
$\bigcirc$ Received important updates from my own locality\\
$\bigcirc$ Received important updates from other regions\\
$\bigcirc$ Received useful updates from Government accounts / NGOs / service providers\\
$\bigcirc$ Received information about safety of others\\
$\bigcirc$ Other \_\_\_\\

\noindent Please describe (in a few sentences) some social media posts that you found to contain important information on Amphan. You can also mention some keywords indicating the type of important posts, or give examples of important posts. \_\_\_\\

\noindent Did social media hinder / worsen the situation related to Amphan in any way? You can select one or more options.

\noindent$\bigcirc$ I did not use social media within 7 days after Amphan\\
$\bigcirc$ I used social media, but did not observe any negative effect of social media\\
$\bigcirc$ Observed fake news / rumours\\
$\bigcirc$ Observed too much useless posts\\
$\bigcirc$ Observed too much of political arguments\\
$\bigcirc$ Observed too much of religious / superstitious posts\\
$\bigcirc$ Observed many posts about other regions, but not enough coverage of my locality\\
$\bigcirc$ Other \_\_\_\\

\noindent Please describe (in a few sentences) some social media posts that you think hindered / worsened the situation.  \_\_\_\\

\noindent \textbf{Preparedness of authorities}

\noindent Were the authorities in your locality (e.g., municipality, electric suppliers, mobile / internet suppliers) prepared to deal with such a disaster situation? Please rate their preparedness in the scale of 1-10, where 1 indicates "Not prepared at all" and 10 indicates "Excellently prepared". \\
1 $\bigcirc$ 2 $\bigcirc$ 3 $\bigcirc$ 4 $\bigcirc$ 5 $\bigcirc$ 6 $\bigcirc$ 7 $\bigcirc$ 8 $\bigcirc$ 9 $\bigcirc$ 10 $\bigcirc$ \\

\noindent When authorities in your locality / helpline numbers were contacted for reporting problems, were the responses helpful?\\
$\bigcirc$ We did not try to contact any local authority / helpline number\\
$\bigcirc$ We tried to contact authorities, but could not reach them / did not get any response\\
$\bigcirc$ We could contact the authorities, but their responses were NOT helpful\\
$\bigcirc$ We could contact the authorities, and received helpful responses\\
$\bigcirc$ Other \_\_\_\\

\noindent If you have any suggestions that will help to better prepare for future disaster events, please list them here. Your suggestions can be general, or specific to your locality. \_\_\_\\

\noindent \textbf{Any other information}

\noindent Please let us know in the text box below if you wish to share any other information about your experiences of cyclone Amphan. E.g., if you wish to give more details about some of your responses in this survey. Or, if you would like to inform about specific social media accounts whose posts you found important or harmful, etc. Specifically, we aim to understand the problems faced after a natural disaster and how digital platforms like social media can help/hinder coping with such disasters. PLEASE CLICK THE "SUBMIT" BUTTON BELOW TO COMPLETE THE SURVEY. \_\_\_\_\_\_\_\_\_\_\_\_\_\_\_\\

\end{scriptsize}

\end{document}